\documentclass[12pt]{article}

\usepackage{latexsym}

\newcommand{\sd}{\mbox{\scriptsize d}}
\newcommand{\Keff}{\ensuremath{K_{\mbox{\scriptsize eff}}}}
\newcommand{\Geff}{\ensuremath{G_{\mbox{\scriptsize eff}}}}
\newcommand{\Zeff}{\ensuremath{Z_{\mbox{\scriptsize eff}}}}
\newcommand{\Geta}{\ensuremath{G(L,C_2^+)}}
\newcommand{\Getamu}{\ensuremath{G(C_2^+,\mu,L)}}
\newcommand{\omtld}{\ensuremath{\tilde{\omega}}}
\newcommand{\dstar}{\ensuremath{\mbox{d}^\star}}
\newcommand{\omp}{\ensuremath{\omega^+}}
\newcommand{\omm}{\ensuremath{\omega^-}}
\newcommand{\ompm}{\ensuremath{\omega^{\pm}}}
\newcommand{\ommp}{\ensuremath{\omega^{\mp}}}
\newcommand{\thp}{\ensuremath{\theta^+}}
\newcommand{\thm}{\ensuremath{\theta^-}}
\newcommand{\thpm}{\ensuremath{\theta^\pm}}
\newcommand{\hhat}{\ensuremath{\hat{h}}}
\newcommand{\khat}{\ensuremath{\hat{K}}}
\newcommand{\Ltld}{\ensuremath{\tilde{\Lambda}}}

\begin{document}

\hfill \vbox{\hbox{UCLA/96/TEP/12}
             \hbox{hep-lat/yymmnn} }

\begin{center}

{\Large \bf SO(3) vortices as a mechanism for generating a mass gap
in the 2d SU(2) principal chiral model 
}\footnote{Research supported in part by NSF Grant PHY89-15286} \\[1cm]

{\bf Tam\'as G.\ Kov\'acs}\footnote{e-mail address: 
kovacs@physics.ucla.edu} \\
{\em Department of Physics \\
University of California, Los Angeles \\
Los Angeles, California 90024-1547} \\[3cm]

\end{center}

\section*{Abstract}

We propose a mechanism that can create a mass gap in the SU(2) chiral spin
model at arbitrarily small temperatures. We give a sufficient condition
for the mass gap to be non-zero in terms of the behaviour of an external
Z(2) flux introduced by twisted boundary conditions. This condition in
turn is transformed into an effective dual Ising model with 
an external magnetic field generated by SO(3) vortices. We show
that having a nonzero magnetic field in the effective Ising
model is sufficient for the SU(2) system to have a mass gap. We also
show that certain vortex correlation inequalities, if satisfied, 
would imply a nonzero effective magnetic field. Finally we give some
plausibility arguments and Monte Carlo evidence for the required correlation
inequalities. 

\vfill

\pagebreak

\section{Introduction}

Two dimensional lattice spin models with a continuous non-Abelian symmetry
are commonly believed to have no phase transition at finite temperature.
According to the Mermin-Wagner theorem in these systems there is no ordered
low temperature phase with spontaneous breakdown of the global symmetry. 
However this does not in itself rule out the presence of a phase transition. 
Indeed the 2d XY model does have a finite temperature Kosterlitz-Thouless
type phase transition characterised by power-law decay of the correlation 
function below the critical temperature as opposed to an exponential 
fall-off in the high temperature phase. Two is the lower critical dimension
for these models and the actual phase structure depends on the properties
of the internal space in a very delicate way. 

In the case of spin models with Abelian symmetries there are rigorous 
results concerning their low temperature behaviour. For discrete symmetries
(e.g.\ the Ising model) there is a distinct low temperature phase with 
long range order while for U(1) symmetry (XY model) the presence of the
KT transition has been rigorously proved \cite{Frohlich}. Recently the exact
mass gap of two dimensional spin models with various different non-Abelian
symmetries has been calculated using the Bethe ansatz \cite{Hasenfratz}.
Scaling of the correlation function has also been observed in a high order
strong coupling expansion of the $N=\infty$ 
SU(N)$\times$SU(N) chiral model \cite{Campostrini}.
It would however be very desirable to have a rigorous proof of a nonzero
mass gap in these models starting from first principles without any 
further assumptions. This would probably also contribute to our intuitive 
understanding of the low temperature behaviour of non-Abelian spin models. 

The situation is possibly even worse in 4d gauge theories. While the U(1) 
theory has been rigorously proved to have a deconfining transition \cite{Guth},
4d non-Abelian gauge theories are believed to remain confining down to 
arbitrarily small couplings but a rigorous proof of this has not yet been
found. In this case we do not even have anything analogous to the Bethe
ansatz solution of two dimensional spin models.
In view of the analogies between 2d spin models and 4d gauge theories
it would be highly desirable to have a unified physical picture of their
low temperature behaviour. In particular we would like to understand 
confinement in gauge theories and the nonzero mass gap in 2d spin models
on a similar ground.

For a few years there has been an ongoing programme to achieve this both in
4d gauge theories \cite{Tomboulis} and in the analogous 2d spin models
\cite{Kovacs,Kovacs1}. In the present paper we report on recent progress
in this direction in the context of the SU(2) principal chiral model.
In this model the degrees of freedom are SU(2)
group elements attached to the sites of a square lattice with a ferromagnetic
nearest neighbour interaction tr$(U_1^\dagger U_2)$. We shall propose a 
mechanism that is sufficient for this system to maintain a mass gap 
at arbitrarily low nonzero temperature. 

Instead of studying the spin-spin correlation function we shall give a 
sufficient condition for the mass gap to be non-zero in terms of the
behaviour of an external Z(2) flux introduced by twisted boundary conditions.
This allows us to explicitly separate the Z(2) degrees of freedom belonging
to the centre of SU(2) and to establish the equivalence of the SU(2) 
spin model with a dual Ising model interacting with SO(3) spins on the
original lattice. We show that a non-zero magnetisation in the dual
Ising model would imply a mass gap in the original SU(2) model. Due to
their coupling to the Ising spins, SO(3) vortices will turn out to 
generate an external magnetic field for the Ising spins, provided
certain vortex correlation inequalities are satisfied. As we shall
see, up to this point all of our arguments are rigorous. Finally we shall 
also give some plausibility arguments and Monte Carlo evidence that the 
vortex correlation inequalities are indeed true. This gives a complete 
picture of the mass gap generation in the SU(2) spin model.

The plan of the paper is as follows. In Section 2 we introduce the order 
parameter that we want to study, the twist. This turns out to be technically
more advantageous than looking at the asymptotic behaviour of the 
spin-spin correlation function and it can be proved that 
``massive'' behaviour of the twist implies the same for the correlation 
function. Also in Section 2 we shall briefly review some properties of
the twist, in particular its high and low temperature expansion and 
its behaviour under duality. In Section 3 we describe the separation of
Z(2) and SO(3) degrees of freedom and discuss the relevant SO(3) vortex
excitations and their dynamics. By an additional Z(2) duality transformation
we map the original SU(2) system on a dual Ising model interacting 
with SO(3) spins. We also rewrite the twist in this language. In Section 4
we develop an effective theory of the dual Ising system by substituting 
the effect of the SO(3) spins with an effective Ising coupling and a magnetic
field. More precisely we derive sufficient conditions for the disorder 
correlation in an effective Ising model to be an upper bound of the twist in 
the original model. These conditions turn out to be vortex correlation
inequalities containing a parameter, the effective magnetic field of
the Ising system. For zero magnetic field they are easily seen to be true
and in Section 5 we argue that a small but nonzero magnetic field can be 
chosen so that the inequalities remain true independently of the
lattice size. Also in Section 5 we present some Monte Carlo evidence
supporting this. In Section 6 we draw our conclusions and make some 
final remarks.

\section{The twist as an order parameter}
\label{sec:twist}

We want to distinguish between a massive and a massless phase in the SU(2) 
spin model. By definition a massive phase is characterised by an exponential
fall-off of the spin-spin correlation function $\langle \frac{1}{2}$tr 
$U_0^\dagger U_x \rangle$ at large distances $x$, 
where $\langle \rangle$ means expectation 
in the infinite volume limit. In our framework it turns out to be technically
more adventageous to consider another operator, a twist winding around the 
lattice and its behaviour as a function of the finite lattice size.\footnote{
The analogous quantity in gauge theories is the sourceless 't Hooft loop 
or magnetic flux free energy that has been extensively used to describe the
phases of lattice gauge theories \cite{Tomb-Yaffe}.} Since the twist
has not been very commonly used as an order parameter in spin models
(but see e.g.\ \cite{twist}), in this section we shall collect some useful
results about it to make the paper self-contained.

\subsection{Notation}

Although we use notations that have become more or less standard in 
(lattice) field theory, we find it useful to include here a brief section
on the notation especially for later reference.

We shall work on a finite two dimensional periodic square lattice  
$\Lambda$. Sites, links and plaquettes of the lattice will be denoted by
$s$, $l$ and $p$. Because of its particular simplicity we shall make use
of the language of group valued forms when writing down the fields and
their interactions. By a $G$-valued $n$-form we mean an assignment of
$G$ elements to oriented elementary $n$-dimensional simplices of the lattice.
Thus e.g.\ a Z(2) valued 2-form $\omega$ assigns a Z(2) element to each 
plaquette and a Z(2) element on a particular plaquette $p$ will be denoted
by $\omega_p$. 

There are two useful operations that can be defined on forms, the exterior
derivative ``d'' and its dual ``\dstar''. They map an $n$-form on an
$n-1$-form (d) and $n+1$-form (\dstar) respectively. By definition
\begin{equation}
 (\mbox{d}\omega)_k = \prod_{m \in \partial k} \omega_m,
\end{equation}
where $k$ denotes a link, $m$ a site if $\omega$ is a 0-form;
$k$ is a plaquette and $m$ is a link if $\omega$ is a 1-form and 
$\partial$ is the boundary operator. Similarly
\begin{equation}
 (\dstar \omega)_m = \prod_{k:m \in \partial k} \omega_k,
\end{equation}
where $m$ is a site, $k$ is a link if $\omega$ is a 1-form and
$m$ is a link, $k$ is a plaquette if $\omega$ is a 2-form. Thus 
e.g.\ the curvature of a Z(2) valued gauge field, a 1-form $A$,
can simply be written as d$A$. For better readability we shall always
omit the parentheses so e.g. d$\omega_m$ means (d$\omega)_m$.

In most of the cases we shall only use Z(2) valued forms the only 
exception being the dynamical variable in the SU(2) chiral model which
is an SU(2) valued 0-form $U_s$ (spins live on lattice sites). The interaction
between the two spins residing at the ends, $s1$ and $s2$ of a given 
link $l$, can be written as d$U_l=\mbox{tr}(U_{s1}^\dagger U_{s2})$. 
This is the only case when the orientation of elementary simplices 
will matter.

When calculating thermal averages involving (discrete or continuous)
group valued fields integration is always understood with respect to the normalised Haar measure.

\subsection{Twist and duality in the Ising model}
     \label{sec:twIsing}

At first we look at the twist in the context of the Ising model since
this will be needed later when we discuss the effective Ising model of
the SU(2) spin system. In the Ising model the twist is not particularly
useful for distinguishing between the phases of the 
theory since the magnetisation provides a much simpler order 
parameter. Nevertheless as we shall see, the twist can also be used for this
purpose.

The partition function of the Ising model is
\begin{equation}
 Z= \prod_s \int d\sigma_s \exp \left( \beta \sum_l \mbox{d}\sigma_l \right)
\end{equation}
with spins $\sigma_s=\pm 1$ living on a finite periodic square lattice. 
Here $\int d\sigma_s$ means integration with the normalised 
invariant measure\footnote{Later we shall also consider 
continuous groups and to make the notations 
uniform we use the same language for discrete groups too.} on Z(2), sites
and links of the lattice are labelled by $s$ and $l$ and finally
d$\sigma_l$ is the product of the two spins on the link $l$.

By definition the twist $\tau_i=\pm 1$ along the direction $i$ is 
the operator that changes the couplings from
$\beta$ to $\tau_i \beta$ on a stack of links winding once around the lattice
along the given direction (Fig \ref{fig:twist}). 
Physically $\tau_i=-1$ means that a topologically 
nontrivial ``domain-wall'' was created along the affected
links. Although this domain-wall is closed, it is not a boundary of any 
region therefore it is impossible to transform it away by a suitable 
redefinition of some of the spin variables. In contrast, if the domain-wall
were a boundary of a region then by a change of variables $\sigma \rightarrow
\tau_i \sigma$ in the region bounded by the twist, it would be possible 
to cancel it and the partition sum would not depend on $\tau_i$.

\begin{figure}[bt!]
\unitlength=1.00mm
\linethickness{0.4pt}
\begin{picture}(114.00,130.00)
\put(10.00,30.00){\line(1,0){100.00}}
\put(110.00,30.00){\line(0,1){100.00}}
\put(110.00,130.00){\line(-1,0){100.00}}
\put(10.00,130.00){\line(0,-1){100.00}}
\linethickness{2.0pt}
\put(55.00,10.00){\line(1,0){10.00}}
\put(69.00,10.00){\makebox(0,0)[lc]{\Large $\eta_l=-1$}}
\put(50.00,40.00){\line(1,0){10.00}}
\put(60.00,50.00){\line(-1,0){10.00}}
\put(50.00,60.00){\line(1,0){10.00}}
\put(60.00,70.00){\line(-1,0){10.00}}
\put(50.00,80.00){\line(1,0){10.00}}
\put(60.00,90.00){\line(-1,0){10.00}}
\put(50.00,100.00){\line(1,0){10.00}}
\put(60.00,110.00){\line(-1,0){10.00}}
\put(50.00,120.00){\line(1,0){10.00}}
\put(60.00,130.00){\line(-1,0){10.00}}
\put(50.00,30.00){\line(1,0){10.00}}
\linethickness{0.4pt}
\put(10.00,70.00){\line(1,0){100.00}}
\put(114.00,70.00){\makebox(0,0)[lc]{{\Large $x'$}}}
\put(6.00,70.00){\makebox(0,0)[rc]{{\Large $x$}}}
\put(90.00,74.00){\makebox(0,0)[cb]{{\Large $C$}}}
\end{picture}
\caption{\small A twist winding around the lattice. 
The links on which the sign of the coupling has been reversed are shown.
\label{fig:twist}}
\end{figure}

By a similar argument it can be easily seen that the twisted stack of 
links can be ``continuously'' deformed by a change of variables and the 
twisted partition sum does not depend on the actual location of 
the twisted links as long as they form a closed loop winding once around 
the lattice.

The two phases of the system can be characterised by its response to such a 
twist. Intuitively one expects that in the high temperature disordered phase,
having a twist does not cost too much energy on a sufficiently large lattice.
On the other hand in the low temperature broken phase where most of the spins
tend to be in one direction, the free energy of the twist is expected 
to grow with the lattice size. 

To make these ideas more precise we can consider the following observables
\begin{equation}
 G(L)= \frac{1}{2} \; \frac{Z_+(L) - Z_-(L)}{Z_+(L) + Z_-(L)} 
 \hspace{10mm} \mbox{and}
 \hspace{10mm} G^{\star}(L)= \frac{Z_-(L)}{\frac{1}{2} \left( 
 Z_-(L) + Z_+(L) \right)},
     \label{eq:defG(L)}
\end{equation}
\begin{equation}
  Z_{\pm}(L)= \int_{\pm} \; d\tau_2 \; Z(\pm 1,\tau_2),
     \label{eq:Zpmdef}
\end{equation}
where $Z(\tau_1,\tau_2)$ is the partition function with twists 
$\tau_1,\tau_2=\pm1$ in the two independent directions and $L$ is the linear
size of the lattice. (The log of) $G^\star(L)$ measures how the free energy
of a twist depends on the lattice size and $G(L)$ is the Z(2)
``Fourier-transform'' of $G^{\star}(L)$. The form of the normalisation
$Z_-(L)+Z_+(L)$ in the denominator was chosen for further convenience, 
we could as well have used just $Z_+(L)$. 

At high temperature (small $\beta$) $G(L)$ decays exponentially with 
increasing lattice size. This can be most easily seen in the high temperature
expansion. Recall that the high temperature expansion of the Ising model
is the sum of closed graphs built of links of the lattice. The weight
of each graph is proportional to $(\tanh \beta)^{\mbox{\tiny \# of links}}$.
In the presence of a twist there is an additional minus sign for each twisted
link contained in the graph. Since the twist 
winds around the lattice, all small loops
will intersect it an even number of times, so these will not contribute
to $G(L)$. The lowest order contribution comes from a loop going all the way
around the lattice in the direction perpendicular to the twist. The shortest 
such loop contains $L$ links and is thus proportional to $(\tanh \beta)^L$.
If $G(L)$ goes exponentially to zero it means that $G^\star(L)$ will go
to 1 in the $L \rightarrow \infty$ limit. This is indeed what we expected,
in the disordered phase the free energy of the twist remains bounded as 
the lattice size goes to infinity.

In the low temperature phase with the Z(2) symmetry broken the behaviour
of $G(L)$ and $G^\star(L)$ is different. Here the free energy cost of 
having a twist grows linearly with the lattice size and $G^\star(L)$ 
goes exponentially to zero while $G(L)$ goes to a nonzero constant
for $L$ large. This is certainly true for the ground 
state and it can also be proved in higher orders of the low temperature
expansion.

Notice that the behaviour of $G(L)$ in the symmetric phase is similar
to that of $G^\star(L)$ in the broken phase and vice versa. This is not 
an accident since $G$ and $G^\star$ are dual to one another in the following
sense. The high temperature expansion of the Ising model with inverse
temperature $\beta$ can be identified
with the low temperature expansion of a dual Ising model with spins
living on the sites of the dual lattice i.e.\ plaquettes of the original
lattice at inverse temperature $\beta^\star$. This gives a mapping between 
the high and the low temperature regimes. It can be proved that the
high temperature expansion of $G(L)$ on the original lattice is
identical to the low temperature expansion of $G^\star(L)$ defined in the dual
system which means that $G(L)_\beta=G^\star(L)_{\beta^\star}$.

Finally we note that in the presence of an external magnetic field the
qualitative behaviour of $G$ and $G^\star$ is the same as in the low 
temperature phase. For sufficiently high temperature the high temperature
expansion is convergent and the leading order exponential fall-off of
$G^\star(L)$ can be rigorously verified.

\subsection{Twist in the SU(2) spin model}
   \label{sec:twistSU(2)}

It is straightforward to generalise the Z(2) twist for any other spin
model with a global symmetry containing Z(2). In principle we could also
use any element of the global symmetry group to define the twist, in spin
models there is no constraint (not like in gauge theories \cite{'tHooft})
restricting the twist to lie in the centre of the symmetry group. In this 
paper however we shall consider only Z(2) twists.

The SU(2)$\times$SU(2) chiral spin model is defined by the partition function
\begin{equation}
 Z = \prod_{s \in \Lambda} \int dU_s
 \hspace{2mm} \exp \left( \beta \sum_{l \in \Lambda}
 \mbox{d}U_l \right),
     \label{eq:PF}
\end{equation}
where the degrees of freedom $U$ are SU(2) group elements on the sites 
of the finite periodic lattice $\Lambda$ and 
d$U_l=\mbox{tr}(U^\dagger_s U_{s'})$ with $[ss']$ being the boundary of the
link $l$. The twists and the quantities $G(L)$ and $G^\star(L)$ can be
defined completely analogously to the Ising model.

The high temperature behaviour of these observables is qualitatively the
same as in the Ising model. At low temperature however there is an important
difference. In the ground state of the twisted Ising system all the spins
are aligned and the energy of the twist is concentrated along a stack 
of links winding around the lattice parallel to the twist. 
On the other hand if the spins can take their values 
in a continuous manifold, the energy of the twist can be spread along 
the direction perpendicular to it.

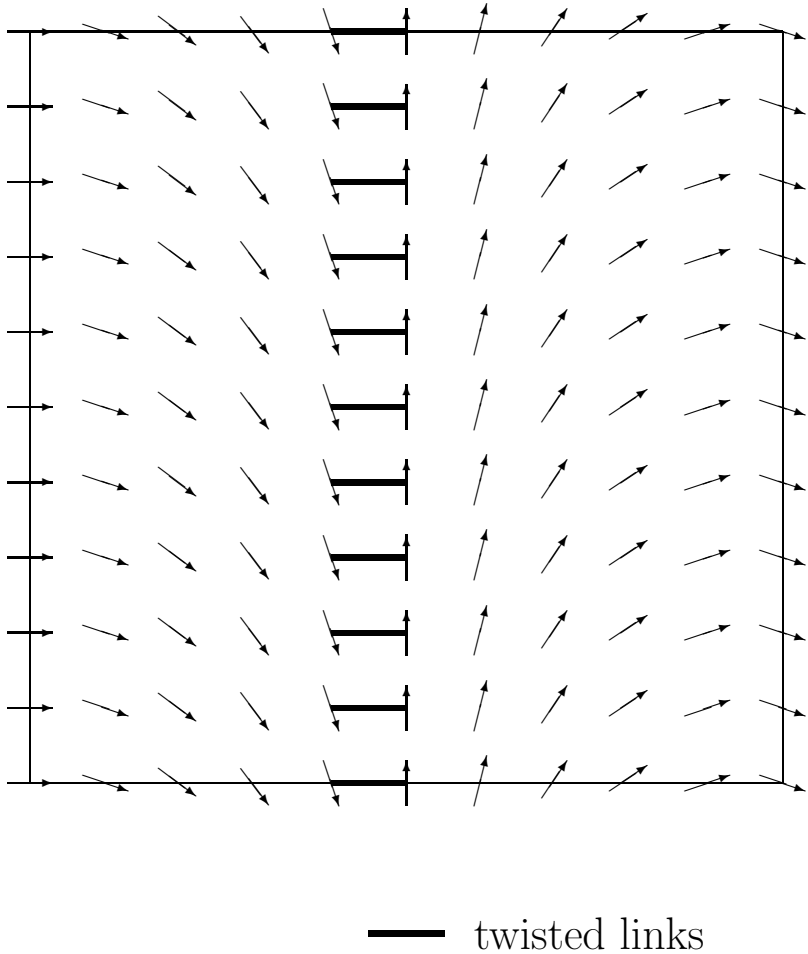
\begin{figure}[tb!]
\unitlength=1.00mm
\linethickness{0.4pt}
\begin{picture}(113.00,130.00)
\put(10.00,30.00){\line(1,0){100.00}}
\put(110.00,30.00){\line(0,1){100.00}}
\put(110.00,130.00){\line(-1,0){100.00}}
\put(10.00,130.00){\line(0,-1){100.00}}
\linethickness{2.0pt}
\put(55.00,10.00){\line(1,0){10.00}}
\put(69.00,10.00){\makebox(0,0)[lc]{\Large twisted links}}
\put(50.00,40.00){\line(1,0){10.00}}
\put(60.00,50.00){\line(-1,0){10.00}}
\put(50.00,60.00){\line(1,0){10.00}}
\put(60.00,70.00){\line(-1,0){10.00}}
\put(50.00,80.00){\line(1,0){10.00}}
\put(60.00,90.00){\line(-1,0){10.00}}
\put(50.00,100.00){\line(1,0){10.00}}
\put(60.00,110.00){\line(-1,0){10.00}}
\put(50.00,120.00){\line(1,0){10.00}}
\put(60.00,130.00){\line(-1,0){10.00}}
\put(50.00,30.00){\line(1,0){10.00}}
\linethickness{0.4pt}
\put(60.00,67.00){\vector(0,1){6.00}}
\put(60.00,77.00){\vector(0,1){6.00}}
\linethickness{2.0pt}
\linethickness{0.4pt}
\put(60.00,27.00){\vector(0,1){6.00}}
\put(60.00,37.00){\vector(0,1){6.00}}
\put(60.00,47.00){\vector(0,1){6.00}}
\put(60.00,57.00){\vector(0,1){6.00}}
\put(60.00,87.00){\vector(0,1){6.00}}
\put(60.00,97.00){\vector(0,1){6.00}}
\put(60.00,107.00){\vector(0,1){6.00}}
\put(60.00,117.00){\vector(0,1){6.00}}
\put(60.00,127.00){\vector(0,1){6.00}}
\put(69.00,127.00){\vector(1,4){1.67}}
\put(78.00,128.00){\vector(2,3){3.33}}
\put(97.00,129.00){\vector(3,1){6.00}}
\put(107.00,131.00){\vector(3,-1){6.00}}
\put(87.00,129.00){\vector(3,2){5.00}}
\put(69.00,117.00){\vector(1,4){1.67}}
\put(69.00,107.00){\vector(1,4){1.67}}
\put(69.00,97.00){\vector(1,4){1.67}}
\put(69.00,87.00){\vector(1,4){1.67}}
\put(69.00,77.00){\vector(1,4){1.67}}
\put(69.00,67.00){\vector(1,4){1.67}}
\put(69.00,57.00){\vector(1,4){1.67}}
\put(69.00,47.00){\vector(1,4){1.67}}
\put(69.00,37.00){\vector(1,4){1.67}}
\put(69.00,27.00){\vector(1,4){1.67}}
\put(78.00,118.00){\vector(2,3){3.33}}
\put(78.00,108.00){\vector(2,3){3.33}}
\put(97.00,119.00){\vector(3,1){6.00}}
\put(107.00,121.00){\vector(3,-1){6.00}}
\put(87.00,119.00){\vector(3,2){5.00}}
\put(97.00,109.00){\vector(3,1){6.00}}
\put(107.00,111.00){\vector(3,-1){6.00}}
\put(87.00,109.00){\vector(3,2){5.00}}
\put(97.00,99.00){\vector(3,1){6.00}}
\put(107.00,101.00){\vector(3,-1){6.00}}
\put(87.00,99.00){\vector(3,2){5.00}}
\put(78.00,98.00){\vector(2,3){3.33}}
\put(78.00,88.00){\vector(2,3){3.33}}
\put(97.00,89.00){\vector(3,1){6.00}}
\put(107.00,91.00){\vector(3,-1){6.00}}
\put(87.00,89.00){\vector(3,2){5.00}}
\put(78.00,78.00){\vector(2,3){3.33}}
\put(78.00,68.00){\vector(2,3){3.33}}
\put(97.00,79.00){\vector(3,1){6.00}}
\put(107.00,81.00){\vector(3,-1){6.00}}
\put(87.00,79.00){\vector(3,2){5.00}}
\put(97.00,69.00){\vector(3,1){6.00}}
\put(107.00,71.00){\vector(3,-1){6.00}}
\put(87.00,69.00){\vector(3,2){5.00}}
\put(97.00,59.00){\vector(3,1){6.00}}
\put(107.00,61.00){\vector(3,-1){6.00}}
\put(87.00,59.00){\vector(3,2){5.00}}
\put(78.00,58.00){\vector(2,3){3.33}}
\put(78.00,48.00){\vector(2,3){3.33}}
\put(78.00,38.00){\vector(2,3){3.33}}
\put(97.00,49.00){\vector(3,1){6.00}}
\put(107.00,51.00){\vector(3,-1){6.00}}
\put(87.00,49.00){\vector(3,2){5.00}}
\put(97.00,39.00){\vector(3,1){6.00}}
\put(107.00,41.00){\vector(3,-1){6.00}}
\put(87.00,39.00){\vector(3,2){5.00}}
\put(97.00,29.00){\vector(3,1){6.00}}
\put(107.00,31.00){\vector(3,-1){6.00}}
\put(87.00,29.00){\vector(3,2){5.00}}
\put(78.00,28.00){\vector(2,3){3.33}}
\put(49.00,133.00){\vector(1,-3){2.00}}
\put(38.00,132.00){\vector(3,-4){3.67}}
\put(27.00,132.00){\vector(4,-3){5.00}}
\put(17.00,131.00){\vector(3,-1){6.00}}
\put(7.00,130.00){\vector(1,0){6.00}}
\put(49.00,123.00){\vector(1,-3){2.00}}
\put(38.00,122.00){\vector(3,-4){3.67}}
\put(27.00,122.00){\vector(4,-3){5.00}}
\put(17.00,121.00){\vector(3,-1){6.00}}
\put(7.00,120.00){\vector(1,0){6.00}}
\put(49.00,113.00){\vector(1,-3){2.00}}
\put(38.00,112.00){\vector(3,-4){3.67}}
\put(27.00,112.00){\vector(4,-3){5.00}}
\put(17.00,111.00){\vector(3,-1){6.00}}
\put(7.00,110.00){\vector(1,0){6.00}}
\put(49.00,103.00){\vector(1,-3){2.00}}
\put(38.00,102.00){\vector(3,-4){3.67}}
\put(27.00,102.00){\vector(4,-3){5.00}}
\put(17.00,101.00){\vector(3,-1){6.00}}
\put(7.00,100.00){\vector(1,0){6.00}}
\put(49.00,93.00){\vector(1,-3){2.00}}
\put(38.00,92.00){\vector(3,-4){3.67}}
\put(27.00,92.00){\vector(4,-3){5.00}}
\put(17.00,91.00){\vector(3,-1){6.00}}
\put(7.00,90.00){\vector(1,0){6.00}}
\put(49.00,83.00){\vector(1,-3){2.00}}
\put(38.00,82.00){\vector(3,-4){3.67}}
\put(27.00,82.00){\vector(4,-3){5.00}}
\put(17.00,81.00){\vector(3,-1){6.00}}
\put(7.00,80.00){\vector(1,0){6.00}}
\put(49.00,73.00){\vector(1,-3){2.00}}
\put(38.00,72.00){\vector(3,-4){3.67}}
\put(27.00,72.00){\vector(4,-3){5.00}}
\put(17.00,71.00){\vector(3,-1){6.00}}
\put(7.00,70.00){\vector(1,0){6.00}}
\put(49.00,63.00){\vector(1,-3){2.00}}
\put(38.00,62.00){\vector(3,-4){3.67}}
\put(27.00,62.00){\vector(4,-3){5.00}}
\put(17.00,61.00){\vector(3,-1){6.00}}
\put(7.00,60.00){\vector(1,0){6.00}}
\put(49.00,53.00){\vector(1,-3){2.00}}
\put(38.00,52.00){\vector(3,-4){3.67}}
\put(27.00,52.00){\vector(4,-3){5.00}}
\put(17.00,51.00){\vector(3,-1){6.00}}
\put(7.00,50.00){\vector(1,0){6.00}}
\put(49.00,43.00){\vector(1,-3){2.00}}
\put(38.00,42.00){\vector(3,-4){3.67}}
\put(27.00,42.00){\vector(4,-3){5.00}}
\put(17.00,41.00){\vector(3,-1){6.00}}
\put(7.00,40.00){\vector(1,0){6.00}}
\put(49.00,33.00){\vector(1,-3){2.00}}
\put(38.00,32.00){\vector(3,-4){3.67}}
\put(27.00,32.00){\vector(4,-3){5.00}}
\put(17.00,31.00){\vector(3,-1){6.00}}
\put(7.00,30.00){\vector(1,0){6.00}}
\end{picture}
\caption{\small The ``semiclassical'' ground state configuration of a
twisted spin model with a continuous spin manifold. The energy of the 
twist can be spread in the direction perpendicular to the twist.
\label{fig:spread}}
\end{figure}

In this case the energetically most favourable configuration has d$U_l<0$
along the twist with the spins roughly antiparallel. The direction of
the spins along the spreading direction changes gradually making half
of a complete turn around the lattice (Fig \ref{fig:spread}). In the 
ground state neighbouring spins in the spreading direction make an angle
proportional to $1/L$ costing an energy $\sim 1/L^2$ and the total 
energy of this configuration goes as $\sim L^2 \times 1/L^2$ on large
lattices. We can see that the energy of a twist does not diverge with the 
lattice size, at least in the classical (zero temperature) approximation.
Intuitively we expect that at nonzero temperature the twist costs even 
less since there is more disorder. This mechanism is analogous to 
flux spreading in gauge theories \cite{Yaffe}.

We saw that both in the Ising model and in the SU(2) spin model
the behaviour of $G(L)$ as a function of the lattice size is 
similar to that of the infinite volume limit of the correlation function.  
This motivates our choice of $G(L)$ as an indicator of the massive phase. 
This can be rigorously justified by proving that in the SU(2) model exponential 
fall-off of $G(L)$ with the lattice size implies the same asymptotic 
behaviour of the infinite volume limit correlation function. (The proof
of the analogous statement in gauge theory \cite{Tomb-Yaffe} can be easily
adapted to the present casa. See \cite{Thesis}.)

\section{Separation of the Z(2) and SO(3) variables}

In the previous section we saw that it is enough to prove the exponential 
decay of the twist $G(L)$ as a function of the lattice size in order to 
establish a mass gap in the SU(2) spin model. Since $G(L)$ is defined in 
terms of a Z(2) twist it will be helpful to disentangle the Z(2) degrees
of freedom belonging to the centre of SU(2) from the rest of the system.
In this Section we shall perform this task by giving a set of exact 
transformations that eventually map the SU(2) model on a dual Ising
model interacting with SO(3) spins on the original lattice. We mainly 
follow \cite{Kovacs} but in addition we also give a more detailed description
of the underlying dynamics of the string and vortex excitations of the 
SO(3) system and their coupling to the Ising spins.

\subsection{Gauging the Z(2) symmetry}

As a first step we introduce new Z(2) valued link variables $A_l$ to gauge
the Z(2) part of the global symmetry. To maintain the equivalence of the
gauged model with the original one the gauge field $A_l$ has to be constrained
to be pure gauge by inserting a delta function $\delta($d$A_p)$ for each
plaquatte. By definition d$A_p$ is the curvature of $A_l$, the product of
link variables around the plaquette $p$ and $\delta(1)=2$, $\delta(-1)=0$.
This is however not enough to make $A_l$ pure gauge. In addition we also 
have to constarain the two independent topologically nontrivial holonomies
$A_{C_i}=\prod_{l \in C_i}A_l$ to be 1. Here $C_1$ and $C_2$ are two loops
built of links winding around the lattice in the two independent directions.

Putting this all together the (nontwisted) partition function reads as
\begin{eqnarray}
 Z(1,1)= \prod_{s \in \Lambda} \int\!\!dU_s \; \prod_{l \in \Lambda} 
         \int\!\!dA_l \; \prod_{p \in \Lambda} \!\! \delta(\mbox{d}A_p) 
         \; \delta(A_{C_1}) \delta(A_{C_2}) \; 
         \exp \left( \beta \sum_{l \in \Lambda} A_l 
         \mbox{d}U_l \right) 
                               \nonumber \\[2mm]
 = \prod_{s \in \Lambda} \int\!\!dU_s \; \prod_{l \in \Lambda} 
   \int\!\!d\sigma_l \; \prod_{p \in \Lambda} 
   \!\!\delta(\mbox{d}\sigma_p \mbox{d}\eta^{-1}_p)
   \; \delta(\sigma_{C_1} \eta^{-1}_{C_1}) 
   \delta(\sigma_{C_2} \eta^{-1}_{C_2}) \; \exp \left( \beta 
   \sum_{l \in \Lambda} \sigma_l |\mbox{d}U_l| \right)
     \label{eq:Zsigma}
\end{eqnarray}
In the second line above we used the notation
\begin{equation}
 \eta_l = \mbox{sign d}U_l = \mbox{sign tr}U^\dagger_{l_1}U_{l_2}
\end{equation} 
and the change of variables $\sigma_l=A_l \eta_l$ to absorb the sign of
d$U_l$ into a redefined gauge field on each link. The variables $\eta_{C_i}$
and $\sigma_{C_i}$ denote the product of $\eta$'s and $\sigma$'s around $C_i$.

The twisted partition sums $Z(\tau_1,\tau_2)$ can also be rewritten in
the same fashion by absorbing the extra signs coming from the twist also in
the $\sigma$'s choosing $\sigma_l=A_l \eta_l \tau_i$ on the twisted links.
The only difference compared to (\ref{eq:Zsigma}) will be that the holonomies
around $C_1$ and $C_2$ will pick up the twist $\tau_2$ and $\tau_1$ along
the direction perpendicular to them. Acoordingly the arguments of the
corresponding delta functions will be multiplied by $\tau_1$ for $C_2$ and
$\tau_2$ for $C_1$. Contractible loops, in particular the boundaries of 
plaquettes always intersect the twists an even number of times therefore 
the plaquette delta functions do not have to be modified.

It is now easy to compute the order parameter $G(L)$ directly from its 
definition
\begin{eqnarray}
 G(L) & = & \frac{\int\!\!d\tau_1 d\tau_2 \; \tau_2 Z(\tau_1,\tau_2)}
             {\int\!\!d\tau_1 d\tau_2 \; Z(\tau_1,\tau_2)}
                       \nonumber \\[2mm]
  & = & \frac{1}{Z} 
  \prod_{s \in \Lambda} \int\!\!dU_s \; \prod_{l \in \Lambda} 
   \int\!\!d\sigma_l \; \prod_{p \in \Lambda} 
   \!\!\delta(\mbox{d}\sigma_p \mbox{d}\eta^{-1}_p)
   \; \sigma_{C_2} \eta^{-1}_{C_2} \; \exp \left( \beta 
   \sum_{l \in \Lambda} \sigma_l |\mbox{d}U_l| \right),
     \label{eq:Gsigma}
\end{eqnarray}
where $Z$ is exactly the same as the numerator of the r.h.s.\ except the
factor $\sigma_{C_2} \eta^{-1}_{C_2}$ is missing. This factor in the numerator
comes from the property of the delta function on Z(2) that
\begin{equation}
 \int\!\!d\tau \; \tau \delta(\tau \sigma) = \sigma.
\end{equation}
The advantage of the normalisation that we chose for $G(L)$ is
that in this way by integrating out the twist we could get rid of the 
delta functions constraining the holonomies around $C_1$ and $C_2$ in 
the ``partition function'' $Z$. This effectively means that we integrated
out the Z(2) part of the boundary conditions.

The remarkable property of this form of the partition function and $G(L)$ is
that the integrand depends on the SU(2) variables only through the 
SU(2)/Z(2) cosets. In other words it has a $U_s \rightarrow -U_s$ local 
Z(2) gauge symmetry and the spins can be regarded SU(2)/Z(2) $\equiv$ SO(3)
variables rather than SU(2) ones. Notice that the $\eta_l$'s themselves are
not gauge invariant but products of them around any closed loop, in
particular around plaquette boundaries and $C_i$ are gauge invariant.
The price that we pay for this extra gauge symmetry is the appearance of
the new Z(2) link variables $\sigma_l$.

\subsection{Z(2) strings and vortices}

At this point it is in order to discuss the physical meaning of the
degrees of freedom and excitations of the system that we obtained by rewriting
the SU(2) model in terms of Z(2) and SO(3) variables.

In (\ref{eq:Gsigma}) there are two different types of couplings between
the SO(3) spins $U_s$ and the Z(2) link variables $\sigma_l$. At first
$\sigma_l$ contributes an extra sign to the coupling between neighbouring
SO(3) spins residing on the two ends of the link $l$. 
If $\sigma_l=1$ then these two
spins are more likely to be parallel whereas for $\sigma_l=-1$ they tend
to be ``perpendicular''. In addition the plaquette delta functions constrain
the ``curvature'' d$\sigma_p=\prod_{l \in \partial p} \sigma_l$ to be 
equal to d$\eta_p$. 

The relevant Z(2) excitations of the model are (stacks of) links with 
$\sigma_l=-1$, which we call $\sigma$-strings. These strings are either
closed or they terminate on plaquettes having d$\sigma_p=-1$ i.e.\ an
odd number of negative $\sigma$'s on their boundary. Due to the plaquette
delta functions the end-plaquettes of $\sigma$-strings have to coincide
with the d$\eta_p=-1$ plaquettes. For reasons that will become clear
later we shall call these vortices. 

In analogy with $\sigma$-strings we can also define $\eta$-strings as the
location of $\eta_l=-1$ links. By construction every vortex is an endpoint
of both a $\sigma$ and an $\eta$-string (Fig \ref{fig:strings}).
Notice however that the location 
of the $\eta$-strings does not have a physical meaning in terms of the 
SO(3) variables, they can be deformed by $U_s \rightarrow -U_s$ gauge
transformations. The location of vortices on the other hand is 
gauge invariant since Z(2) gauge transformations always change the sign of
an even number of factors in d$\eta_p$. 
The situation is analogous to gauge theories 
where one has Dirac strings attached to monopoles and the strings can 
be deformed by gauge transformations but their endpoints (boundaries in 4d), 
the monopoles (monopole loops) are gauge invariant objects.

\begin{figure}[tb!]
\unitlength=1.00mm
\linethickness{0.4pt}
\begin{picture}(110.00,130.00)
\put(10.00,30.00){\line(1,0){100.00}}
\put(110.00,30.00){\line(0,1){100.00}}
\put(110.00,130.00){\line(-1,0){100.00}}
\put(10.00,130.00){\line(0,-1){100.00}}
\put(50.00,60.00){\line(-1,0){10.00}}
\put(40.00,70.00){\line(1,0){10.00}}
\put(50.00,80.00){\line(-1,0){10.00}}
\put(40.00,90.00){\line(1,0){10.00}}
\put(45.00,95.00){\makebox(0,0)[cc]{\Large $\otimes$}}
\put(45.00,45.00){\makebox(0,0)[cc]{\Large $\otimes$}}
\put(50.00,50.00){\line(-1,0){10.00}}
\linethickness{2.0pt}
\put(50.00,50.00){\line(0,-1){10.00}}
\put(60.00,40.00){\line(0,1){10.00}}
\put(70.00,50.00){\line(0,-1){10.00}}
\put(70.00,50.00){\line(1,0){10.00}}
\put(80.00,60.00){\line(-1,0){10.00}}
\put(70.00,70.00){\line(1,0){10.00}}
\put(80.00,70.00){\line(0,1){10.00}}
\put(80.00,80.00){\line(1,0){10.00}}
\put(90.00,90.00){\line(-1,0){10.00}}
\put(80.00,90.00){\line(0,1){10.00}}
\put(70.00,100.00){\line(0,-1){10.00}}
\put(60.00,90.00){\line(0,1){10.00}}
\put(50.00,100.00){\line(0,-1){10.00}}
\put(10.00,10.00){\makebox(0,0)[lc]{\Large $\otimes$ vortex}}
\linethickness{0.4pt}
\put(40.00,10.00){\line(1,0){10.00}}
\put(54.00,10.00){\makebox(0,0)[lc]{\Large $\sigma_l=-1$}}
\linethickness{2.0pt}
\put(85.00,10.00){\line(1,0){10.00}}
\put(99.00,10.00){\makebox(0,0)[lc]{\Large $\eta_l=-1$}}
\end{picture}
\caption{\small A typical configuration with one vortex pair and the 
$\eta$ and $\sigma$ string connecting them.  
\label{fig:strings}}
\end{figure}

It should be noted that there is a substantial difference between the 
energetics of the $\sigma$ and $\eta$ strings. To see this let us look at
a closed contour of links enclosing exactly one vortex. Both the $\sigma$ and
the $\eta$ string attached to the vortex have to pierce the contour somewhere.
The link $l$ where the $\sigma$ string intersects the contour is unambiguously
given by the $[\sigma]$ configuration and it carries an energy 
$\sim 2\,$d$U_l$. On the other hand as we have already seen, 
the location $l'$ where the 
$\eta$ string crosses the contour is not Z(2) gauge invariant.
It is thus not surprising that $l'$ has no distinguished role
among the links of the contour; it does not carry any extra energy.
Indeed in a typical low temperature SO(3) configuration the spins vary 
slowly around the contour and they make half of a compelete turn as the
contour encircles the vortex. This is more favourable both in terms of energy
and entropy than having abrupt changes. Now we can see why the d$\eta_p=-1$
plaquettes were called vortices. d$\eta_p$ is well defined in terms of the
SO(3) variables and a ``smooth'' vortex configuration is a special case 
of defects appearing in 2d spin models due to the non simply-connectedness of
the internal space \cite{Mermin}. 

At this point we want to emphasize that the internal space of the original 
SU(2) spin model i.e.\ the SU(2) group manifold is simply connected,
therefore no vortices are present in the original model. After separating
the Z(2) degrees of freedom however, the remaining configuration space
(at each site) is essentially SO(3). Due to the doubly connected nature of
SO(3) this representation admits vortices characterised by a Z(2) charge.
There is no contradiction between the two representations of the
model because the SO(3) vortices always come along with d$\sigma_p=-1$
defects in the Z(2) part of the system (this is ensured by the constraints
in eq.\ (\ref{eq:Zsigma})). Were we to rewrite the model again in the 
original SU(2) language the two types of defects, the vortices and 
the Z(2) defects in $\sigma$, would always combine to give defects in the
SU(2) language that are topologically trival.

Let us now consider two vortices connected by an $\eta$ and a $\sigma$ string.
It is clear that at low temperature the interaction between the two 
vortices is dominated by the $\sigma$ string since its energy 
is proportional to its length. On the other hand from
a simple semiclassical estimate (i.e.\ using a slowly varying ``smooth'' SO(3)
configuration) the energy of the $\eta$ string can be easily seen to go
as the log of its length. At low temperatures the vortices are in closely
bound pairs due to the linearly confining potential of the 
$\sigma$ string.  Vortex pairs with short $\sigma$ strings between 
them are local excitations, they are not expected to
have a big influence on the large scale behaviour of the system. 
The $\eta$ strings however can still fluctuate considerably 
because of their smaller energy cost and as we shall see long $\eta$ strings
are capable of disordering the system even on large distance scales.

At first sight one would expect that at sufficiently low temperature
due to their logarithmic behaviour even the long $\eta$ strings will
freeze out of the system. This is however not quite right. To see this 
we have to notice that in the above semiclassical energy estimate
we assumed that the lattice is very large compared to the length of the
$\eta$ string. Since we shall not be working directly in the infinite
volume limit but rather consider the lattice size dependence of $G(L)$,
this assumption may not be right. Indeed, let us look at 
two nearby vortices with a short $\sigma$ string in between but
the $\eta$ string connecting them going all the way around the lattice.
By a semiclassical argument similar to the one in Section 
\ref{sec:twistSU(2)} we can see that the energy of this configuration
goes to a constant when $L \rightarrow \infty$. The two vortices of course
can be removed by closing the $\eta$ string around the lattice.

These types of configurations can potentially have a long range disordering
effect. In fact it is exactly these configurations that have to be present
with a sufficient weight to ensure that the relative difference between 
the  twisted partition sum and the untwisted one decreases sufficiently 
rapidly with the lattice becoming larger. As we shall see later, 
this is what we need for $G(L)$ to decay exponentially.

\subsection{Duality transformation}

In the previous section we have succeded in separating Z(2) and SO(3) 
variables in the SU(2) spin model and showed 
that vortex and string excitations of
the SO(3) part can have a long range disordering effect. Unfortunately 
the Z(2) part of the system does not look very familiar and in order to
proceed we want to cast it into a more managable form. This can be done
by a duality transformation on the $\sigma$ link variables which amounts
to trading them for Z(2) plaquette variables that we call $\omega_p$. 
The interaction between $\omega$'s will turn out to be nearest neighbour
ferromagnetic although with fluctuating couplings that depend on the SO(3)
spin configuration. The $\omega$'s in this way give an Ising model on
the dual lattice the sites of which are plaquettes of the original lattice.

Technically the Z(2) duality transformation is done by expanding each factor
in the integrand of $G(L)$ (equation (\ref{eq:Gsigma})) depending on the
$\sigma$'s in terms of Z(2) characters as
\begin{equation}
 f(\sigma) = \hat{f}(1) + \sigma \hat{f}(-1) = 
 \int \!\! d\alpha \, \hat{f}(\alpha) \; \chi_\alpha(\sigma).
     \label{eq:chexp}
\end{equation}
The expansion for the different factors in (\ref{eq:Gsigma}) reads as
\begin{eqnarray}
 \delta(\mbox{d}\sigma_p \mbox{d}\eta^{-1}_p) & =  &
 \int \!\! d\omega_p \; \chi_{\omega_p}(\mbox{d}\sigma_p \mbox{d}\eta^{-1}_p)
                 \nonumber \\
 \sigma_{C_2} & = & \chi_{-1}(\sigma_{C_2})  =  
 \prod_{l \in C_2} \chi_{-1}(\sigma_l)
                 \nonumber \\
 e^{\sigma_l |\sd U_l|} & = & 
 \int \!\! d\alpha_l \; \chi_{\alpha_l}(\sigma_l) 
 \, e^{\mathcal{L}(\sd U_l,\alpha_l)},
\end{eqnarray}
where $e^{\mathcal{L}(\sd U_l,\alpha_l)}$ is the Z(2) ``Fourier transform''
of $e^{\sigma_l |\sd U_l|}$ and for later convenience it can be split
into a part depending only on the SO(3) variables and another one depending
both on the SO(3) and Z(2) degrees of freedom;
\begin{equation}
 e^{\mathcal{L}(\sd U_l,\alpha_l)} = \frac{1}{2} 
 \left( e^{\beta |\sd U_l|} + \alpha_l \, e^{-\beta |\sd U_l|} \right)
 = e^{ M(\sd U_l)} \; e^{ \alpha_l K(\sd U_l)}.
\end{equation}
Here the functions $K$ and $M$ are given by
\begin{equation}
 M(\mbox{d}U_l) = \frac{1}{2} \ln \sinh(2 \beta |\mbox{d}U_l|) \hspace{1cm}
 \mbox{and} \hspace{1cm} K(\mbox{d}U_l) = 
 \frac{1}{2} \ln \coth(\beta |\mbox{d}U_l|).
     \label{eq:KM}
\end{equation}
Substituting these into the expression (\ref{eq:Gsigma}) of $G(L)$ the 
$\sigma$ variables can be explicitly integrated out using the orthogonality
of the characters. This will give rise to constraints between the remaining
Z(2) degrees of freedom, namely on each link 
\begin{equation}
 \alpha_l = \dstar \omega_l = \prod_{p:l \in \partial p} \omega_p
\end{equation}
except on the links belonging to $C_2$ where 
$\alpha_l= -\dstar \omega_l$. These constraints make the $\alpha_l$ integrals
trivial and yield 
\begin{eqnarray}
 \lefteqn{   G(L) = \frac{1}{Z} \int\!\!d\nu[U]\, 
 \prod_{p \in \Lambda} \int\!\!d\omega_p \chi_{\omega_p}(\mbox{d}\eta_p) \;
    }         \hspace{3cm}      
                        \nonumber   \\[2mm]
 & & \times \eta_{C_2} \exp \left( \sum_{l \notin C_2} K(\mbox{d}U_l) \, 
 \dstar \omega_l  \; - \sum_{l \in C_2} 
  K(\mbox{d}U_l) \, \dstar \omega_l \right),
     \label{eq:Gomega}
\end{eqnarray}
where
\begin{equation}
 d \nu[U] = \prod_{s \in \Lambda} d U_s \, e^{\beta \sum_l M(\sd U_l)}
     \label{eq:SO(3)measure}
\end{equation}
is the SO(3) (Z(2) gauge invariant) part of the measure.

Exactly the same transformation can be carried out on the partition 
function $Z$, the only difference in the final result compared to 
(\ref{eq:Gomega}) will be the absence of the factor $\eta_{C_2}$ and
the minus sign in the couplings between $\omega$'s along $C_2$.

In this representation the system consists of Z(2) spins $\omega_p$ attached
to plaquettes and the SO(3) part of the system is unchanged.
 The $\omega$'s on nearest neighbour plaquettes
sharing the link $l$ interact via the coupling $K(\mbox{d}U_l)$. The 
Z(2) part of the system is essentially an Ising model on the dual lattice
although the spin-spin couplings can fluctuate since they depend on the 
SO(3) configuration.

For large $\beta$ the effective SO(3) action $M($d$U_l)$ is peaked at the 
maximum of $|$d$U_l|$ which means that $\beta |$d$U_l|$ is also large for
``most'' of the relevant configurations. The asymptotic behaviour of 
$M($d$U_l)$ in this limit is $\sim \beta |$d$U_l|-\ln2\,/2$. On the other 
hand the duality transformation changes low temperature to high temperature
for the Z(2) part of the system. Indeed the $\beta \rightarrow \infty$ limit
of the effective Ising coupling is $K($d$U_l) \sim e^{-2 \beta |\sd U_l|}$.

\section{Construction of the effective Ising model}
\label{sec:effIs}

We have seen that after separating the Z(2) and SO(3) variables in the 
SU(2) model, $G(L)$ can be rewritten in terms of a dual Ising model
coupled to SO(3) spins. As can be seen from equation (\ref{eq:Gomega})
the Ising part of $G(L)$ is essentially the ratio between 
a twisted (along $C_2$) and the untwisted partition sum, i.e.\ the
free energy of a twist along $C_2$ ($G^\star(L)$ in the language of 
Section \ref{sec:twist}). Notice that because of the duality transformation,
$G(L)$ in the original model is  becomes $G^\star(L)$ in the dual Ising
model.

We have already discussed that the Ising couplings between nearest neighbour
plaquettes depend on the SO(3) spin configuration. 
Besides there are two more SO(3) dependent 
pieces in the expression of $G(L)$; $\eta_{C_2}$ and the product
of group characters $\chi_{\omega_p}(\mbox{d}\eta_p)$. While $\eta_{C_2}$
depends solely on the SO(3) variables, the group characters couple vortices 
to the Ising spins on each plaquette. 

For a fixed SO(3) configuration with vortices at $(p_1,p_2...p_{2n})$
and $S$ $\eta$ strings\footnote{Although $S$ is not well 
defined in terms of the SO(3) variables, its parity $(-1)^S$ is invariant
under coset reparametrisations} crossing $C_2$ these give an overall factor 
of $(-1)^S \prod_{i=1}^{2n} \omega_{p_i}$. Were it not for this additional 
factor depending on the SO(3) configuration, the Z(2) part of the system
would be a ferromagnetic Ising model at high temperature with unbroken 
symmetry and $G(L)$, the ratio of the twisted and untwisted partition 
sum would go to a nonzero constant on large lattices. It follows that
if we were to eliminate all the vortices from the measure and also constrain
$\eta_{C_2}$ to be +1 the exponentially falling asymptotic behaviour of
$G(L)$ would be lost. The configurations responsible for the correct asymptotic
behaviour of $G(L)$ are exactly those that would be eliminated by the above
constraints.

\subsection{Strings and vortices in the dual representation}

Let us now have a closer look at the relevant configurations in the different
vortex sectors. In the absence of vortices the only difference between
the numerator and the denominator of (\ref{eq:Gomega}) is the factor 
$\eta_{C_2}$. This is true up to graphs of size $\sim L$ 
in the high temperature expansion of the $\omega$ Ising model 
but these are exponentially suppressed (see Section \ref{sec:twIsing}).
All the short contractibe $\eta$ strings have to cross
$C_2$ an even number of times and they do not contribute to $\eta_{C_2}$.
$\eta_{C_2}=-1$ thus means that there is an odd number of $\eta$ strings 
winding around the lattice perpendicularly to $C_2$. 
In the zero-vortex sector it is exactly the long topologically nontrivial
$\eta$ strings that give a negative contribution to the numerator of 
$G(L)$ and a positive one to the denominator. If there were no vortices
at all, the asymptotic behaviour of $G(L)$ would solely depend on the 
relative weigth of these configurations.

\begin{figure}[tb!]
\unitlength=1.00mm
\linethickness{0.4pt}
\begin{picture}(110.00,130.00)
\put(10.00,30.00){\line(1,0){100.00}}
\put(110.00,30.00){\line(0,1){100.00}}
\put(110.00,130.00){\line(-1,0){100.00}}
\put(10.00,130.00){\line(0,-1){100.00}}

\linethickness{1.0pt}
\put(80.00,25.00){\line(1,0){10.00}}
\put(94.00,25.00){\makebox(0,0)[lc]{\Large $\eta_l=-1$}}
\put(50.00,60.00){\line(1,0){10.00}}
\put(50.00,80.00){\line(1,0){10.00}}
\put(50.00,30.00){\line(1,0){10.00}}

\put(60.00,60.00){\line(0,-1){10.00}}
\put(70.00,50.00){\line(0,1){10.00}}
\put(70.00,60.00){\line(1,0){10.00}}
\put(80.00,69.00){\line(-1,0){10.00}}
\put(70.00,80.00){\line(1,0){10.00}}
\put(80.00,90.00){\line(-1,0){10.00}}
\put(70.00,90.00){\line(0,1){10.00}}
\put(60.00,100.00){\line(0,-1){10.00}}
\put(60.00,90.00){\line(-1,0){10.00}}

\linethickness{0.4pt}
\put(10.00,70.00){\line(1,0){100.00}}
\put(90.00,74.00){\makebox(0,0)[cb]{{\Large $C_2$}}}
\put(55.00,75.00){\makebox(0,0)[cc]{{\Large $\otimes$}}}
\put(55.00,65.00){\makebox(0,0)[cc]{{\Large $\otimes$}}}
\put(25.00,25.00){\makebox(0,0)[lc]{{\Large $\otimes$ vortex}}}
\linethickness{4pt}
\put(50.00,70.00){\line(1,0){10.00}}
\put(40,5){\line(1,0){10.00}}
\put(70,5){\makebox(0,0)[cc]{{\Large $\omega$-string}}}
\end{picture}
\caption{\small A vortex pair with the connecting $\eta$ and $\omega$ string 
forming a contractible loop. This graph gives the same contribution to 
the numerator and the denominator of $G(L)$. \label{fig:gaploopb}}
\end{figure}
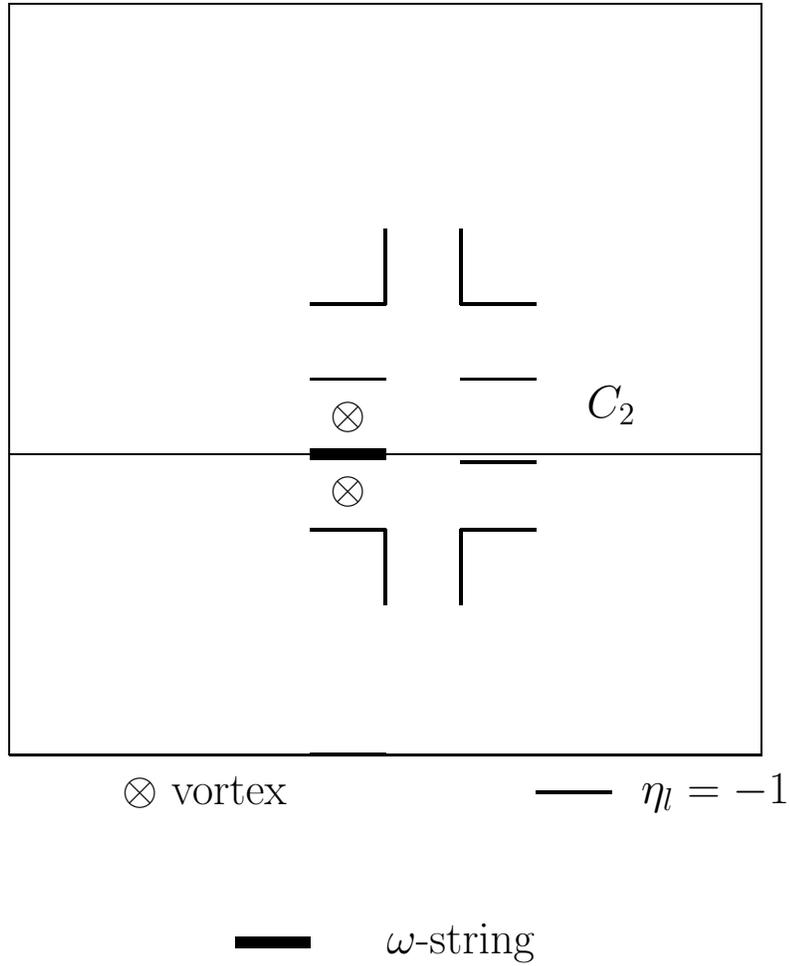

Let us now consider the sector with two nearby vortices at $p_1$ and $p_2$.
In this sector the high-temperature expansion of the $\omega$ spin system
contains small graphs connecting $p_1$ and $p_2$. If a graph like this crosses
$C_2$ then in the numerator it acquires a minus sign due to the twist.
Besides we still have the factor $\eta_{C_2}$. The combined effect
of these two signs will be different for the numerator and the denominator
of (\ref{eq:Gomega}) only if the $\omega$ graph together with the $\eta$
string connecting the two vortices close into a noncontractible loop
(Fig \ref{fig:gaploopa}). This is absolutely essential, otherwise the
sign coming from $\eta_{C_2}$ would cancel the one coming from the twist.
In other words the curve $C_2$ could be deformed by a Z(2) gauge 
transformation to decouple from both the $\omega$ and the $\eta$ string
(as in Fig \ref{fig:gaploopb}). This obviously cannot be done with 
the configuration in Fig \ref{fig:gaploopa}.

\begin{figure}[tb!]
\unitlength=1.00mm
\linethickness{0.4pt}
\begin{picture}(110.00,130.00)
\put(10.00,30.00){\line(1,0){100.00}}
\put(110.00,30.00){\line(0,1){100.00}}
\put(110.00,130.00){\line(-1,0){100.00}}
\put(10.00,130.00){\line(0,-1){100.00}}
\linethickness{1.0pt}
\put(80.00,20.00){\line(1,0){10.00}}
\put(94.00,20.00){\makebox(0,0)[lc]{\Large $\eta_l=-1$}}
\put(50.00,40.00){\line(1,0){10.00}}
\put(60.00,50.00){\line(-1,0){10.00}}
\put(50.00,60.00){\line(1,0){10.00}}
\put(50.00,80.00){\line(1,0){10.00}}
\put(60.00,90.00){\line(-1,0){10.00}}
\put(50.00,100.00){\line(1,0){10.00}}
\put(60.00,110.00){\line(-1,0){10.00}}
\put(50.00,120.00){\line(1,0){10.00}}
\put(60.00,130.00){\line(-1,0){10.00}}
\put(50.00,30.00){\line(1,0){10.00}}
\linethickness{0.4pt}
\put(10.00,70.00){\line(1,0){100.00}}
\put(90.00,74.00){\makebox(0,0)[cb]{{\Large $C_2$}}}
\put(55.00,75.00){\makebox(0,0)[cc]{{\Large $\otimes$}}}
\put(55.00,65.00){\makebox(0,0)[cc]{{\Large $\otimes$}}}
\put(25.00,20.00){\makebox(0,0)[lc]{{\Large $\otimes$ vortex}}}
\linethickness{4pt}
\put(50.00,70.00){\line(1,0){10.00}}
\put(40,5){\line(1,0){10.00}}
\put(70,5){\makebox(0,0)[cc]{{\Large $\omega$-string}}}
\end{picture}
\caption{\small A vortex pair with its connecting $\eta$ and $\omega$ string
forming a noncontractible loop winding around the lattice. 
This configuration gives different contributions to the numerator and
denominator of $G(L)$.
\label{fig:gaploopa}}
\end{figure}
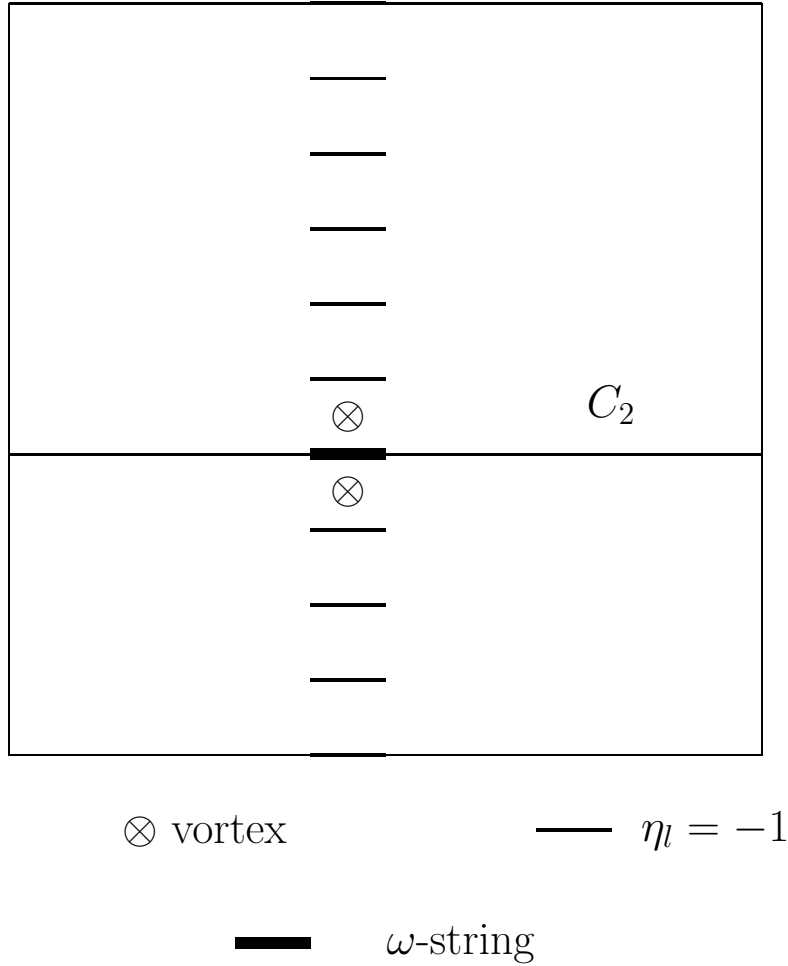

The configuration in Fig \ref{fig:gaploopa} is not very much different
from the one discussed in the 0-vortex sector. Again it contains a long
$\eta$ string winding almost around the lattice connecting the two
vortices with a small gap between them.

We can already see the similarity between this expansion and the high
temperature expansion of the Ising model with a nonzero magnetic field
discussed in Section \ref{sec:twIsing}. The only difference here is that 
the graphs of order $2n$ are proportional to $2n$-vortex expectations
instead of the $2n$-th power of the magnetic field. It should be stressed
that the vortex expectations have to be calculated with the SO(3) measure
(\ref{eq:SO(3)measure}) which is ``blind'' to the centre of SU(2). 

The similarity between the two expansions hints that it might be possible
to approximate the dual representation of the SU(2) model with an
effective Ising model. In this
way we could trade the SO(3) variables for a suitably chosen
effective coupling \Keff$(\beta)$ and an effective
magnetic field $h(\beta)$. To put it more precisely, we want to bound
expression (\ref{eq:Gomega}) of $G(L)$ by the ratio of the twisted and
untwisted partition function of an Ising model with a suitably chosen
coupling \Keff$(\beta)$ and a magnetic field $h(\beta)$. For large 
$\beta$ \Keff\ and $h$ will be exponentially small and the high temperature
expansion for the effective Ising model will be convergent. As we saw in
Section \ref{sec:twIsing} in the presence of a nonzero magnetic field
the ratio of the twisted partition function to the untwisted one falls off
exponentially with the lattice size. It means that if the Ising bound 
for $G(L)$ can be established with a {\em nonzero} magnetic field it would
imply the exponential decay of $G(L)$. 

To summarize, in order to explicitly construct the effective Ising model
we have to deal with three questions.

\begin{itemize}

\item{The fluctuating couplings $K($d$U_l)$ have to be approximated by some
\Keff.}
\item{The effect of the SO(3) vortices on the Ising spins has to be 
substituted with an external magnetic field.}
\item{Somehow we have to get rid of the factor $\eta_{C_2}$ that depends
only on the SO(3) variables.}

\end{itemize}

The last point turns out to be the simplest one. Since long $\eta$ strings
disorder the system, constraining $\eta_{C_2}$ to be $+1$ makes the system
more ordered and increases $G(L)$. Indeed by using reflection positivity
it can be proved that this constraint increases $G(L)$ \cite{Thesis}.
From this point on we shall consider the quantity \Geta\ which is
defined as $G(L)$ calculated with this additional constraint. The SO(3) 
measure $d\nu[U]$ supplemented with this constraint will be called 
$d\nu_+[U]$.

\subsection{The effective Ising coupling}

We now want to compare $\Geta$ with the quantity
\begin{equation}
 \Geff(h,\Keff,L) = \frac{1}{\Zeff} \prod_{s \in \Lambda} \int\!\! d\omega_p
 \exp \left( \Keff \sum_{l \notin C_2} \dstar \omega_l 
 - \Keff \sum_{l \in C_2} \dstar \omega_l + h 
 \sum_{p \in \Lambda} \omega_p \right),
     \label{eq:Geff}
\end{equation}
i.e.\ the ratio of the twisted and untwisted partition function of
an Ising model. 

In the original SU(2) model we started with the particular 
form of the interaction $\mbox{d} U_l=\mbox{tr}U_{1l}^\dagger U_{2l}$.
We could have as well used any other action in the same universality 
class. In particular the choice
\begin{equation}
 S = \sum_{l \in \Lambda} |\mbox{d} U_l | + \mu \sum_{l \in \Lambda} \eta_l
     \label{eq:Snew}
\end{equation}
with $\mu>0$ would have considerably simplified the discussion. After
the separation of the Z(2) variables and the duality transformation
on this action we get
\begin{equation}
 M(\mbox{d} U_l) = \beta |\mbox{d} U_l | \hspace{1cm} \mbox{and} \hspace{1cm}
 K(\mbox{d} U_l) = \frac{1}{2} \ln \coth 2 \mu \beta,
     \label{eq:KMnew}
\end{equation}
i.e.\ the Ising coupling is independent of the SO(3) variables and we do
not need to find an approximation for \Keff. Because of its 
simplicity we shall only consider this form of the action and the quantity
$\Getamu$ corresponding to $\Geta$ with this particular choice of the
SU(2) action.

\subsection{Ginibre decomposition}

We now want to find a nonzero magnetic field $h$,
independent of the lattice size such that on any lattice
\begin{equation}
 \Geff(h,K,L) - \Getamu \geq 0.
\end{equation}
This would imply the exponential fall-off of \Getamu\ and by the inequalities
proved so far the same behaviour for the correlation function of the SU(2)
spin model.

By an argument similar to Ginibre's proof
of the Griffiths' inequalities \cite{Ginibre} the difference 
$\Zeff Z (\Geff - G)$ can
be decomposed into a sum of terms independent of the $\omega$ variables,
containing only vortex correlations calculated with the SO(3) measure
$d\nu_+[U]$, including the constraint on the $\eta$ string.
Details of this decomposition can be found in the Appendix. 
The form of these terms is (up to non-negative constant factors)
\begin{equation}
 D(\hhat) = \langle \prod_{p \in \bar{\Lambda}} (\thm_p \pm \hhat \thp_p)
 \prod_{p \notin \bar{\Lambda}} (\thp_p \pm \hhat \thm_p) \rangle_L,
     \label{eq:vorex}
\end{equation}
where $\langle \rangle_L$ means expectation with respect to the SO(3) measure
$d\nu_+[U]$, $\bar{\Lambda}$ is some subset of $\Lambda$ containing an even
number of plaquettes and $\hhat= \tanh h$.

It is easily seen that for $\hhat=0$ all these diagrams are strictly positive
implying that $\Geff(h=0,K,L) > \Getamu$. On any finite lattice \Geff\
is an analytic function of $h$ and this inequality is in fact true not only
for $h=0$ but for any $h$ in some finite interval around 0. However we still
need to establish that this  interval does not shrink to
zero as the lattice size goes to infinity.

\section{Vortex correlations}

In this section we want to study the dependence of the vortex
expectations  (\ref{eq:vorex}) on the effective magnetic field 
and make it plausible that there is a finite interval 
around $h=0$ for which all these expectations are non-negative 
independently of the lattice size. We would like to emphasize that
up to this point all our arguments were rigorous.

\subsection{Factorisation inequality}

Vortex expectations of the type (\ref{eq:vorex}) come in huge varieties.
We can introduce some order into this abundance by explicitly
constraining out some of the Z(2) excitations while still keeping
a finite density of them. This will certainly make the system more
ordered and the presence of a mass gap in this system would
imply a mass gap in the original model. With these further constraints
we can essentially pair up vortices so that the remaining nonzero 
expectations of the type (\ref{eq:vorex}) will have the form
\begin{equation}
 D(\hhat) = \langle \prod_{P_i \in \Ltld} (\thm_{P_i}- \hhat \thp_{P_i}) 
 \; \prod_{p \notin \Ltld} \thp_p \rangle,
     \label{eq:modvorex}
\end{equation}
where $\Ltld$ is a sublattice of $\Lambda$, $P_i=\{ p_{i1} p_{i2}\}$
denotes pairs of plaquettes and $\thpm_{P_i}= \thpm_{p_{i1}} \thpm_{p_{i2}}$.

Now for $\hhat=0$ it can be proved that 
\begin{equation}
 \langle \prod_{P_i \in \Ltld} \thm_{P_i} \prod_{p \notin \Ltld}
 \thp_p \rangle \; \geq \; 
 \prod_{i=1}^n \langle \thm_{P_i} \prod_{p \notin P_i} \thp_p \rangle.
     \label{eq:facineq}
\end{equation}
This correlation inequality can be checked by using 
reflection positivity for vortex configurations that are
symmetrically placed about a line bisecting the lattice.
For more general configurations the proof involves
the application of the FKG inequalities \cite{FKG}.

Now it is quite plausible that if the above factorisation inequality holds for
$\hhat=0$ then it will be true for some finite interval around $\hhat=0$
which is independent of the lattice size. Let us for the moment assume this. 
It follows then that we have
a lower bound on $\Geff - G$ in terms of a sum that contains products of
two-vortex expectations of the form
\begin{equation}
 \prod_{i=1}^n \langle (\thm_{P_i} -
 \hhat^2 \thp_{P_i}) \prod_{p \notin P_i} \thp_p \rangle
\end{equation}
It is now enough to establish that all these 
factors are non-negative for any $\hhat$ in some finite interval around
zero independent of the lattice size. This is equivalent to the 
statement that the free energy of a vortex pair (as compared to that 
of the no-vortex state),
\begin{equation}
 F_L(p_1,p_2) = -\frac{1}{\beta} \frac{\langle \thm_{p1} \thm_{p2}
 \prod_{p \neq p1, p2} \thp_p \rangle}{\langle \prod_{p \in \Lambda} 
 \thp_p \rangle},
\end{equation}
is bounded for arbitrarily large lattices. Normally one would think about
a pair of vortices as a local excitation and its free energy is not expected
to diverge with the lattice size. This is however not trivially true
in our case due to the constraint on the $\eta$ sting contained in the
measure $d\nu_+[U]$. If the two nearby
vortices are cut off from each other by $C_2$, their connecting $\eta$ string
has to wind around the lattice to avoid crossing $C_2$. This type of excitation
is definitely not a local object. In fact the ``most non-local'' excitation
in the two-vortex sector is the one that has two adjecent vortices separated
by $C_2$ (see Figure \ref{fig:gaploopa}). All other types of 
two-vortex configurations contain shorter
$\eta$ strings and therefore have a smaller free energy. 

In two dimensions, as we have already seen, the free energy of such
a long $\eta$ string stays finite as $L \rightarrow \infty$ in the
semiclassical approximation. This is due to flux spreading that allows the
cost of the string creation to spread laterally in the direction perpendicular
to the string. Intuitively one expects the spreading of the flux to be even
faster than that given by the semiclassical approximation but unfortunately
we could not prove this analytically. Instead we measured the
lattice size dependence of the two-vortex free energy using Monte Carlo.

\subsection{Monte Carlo results}

In this subsection we want to give some evidence that the free energy of
two adjecent vortices separated by the curve $C_2$ does not diverge when the
lattice size goes to infinity. Moreover we claim that this is very likely to
be true for any non-zero temperature. 

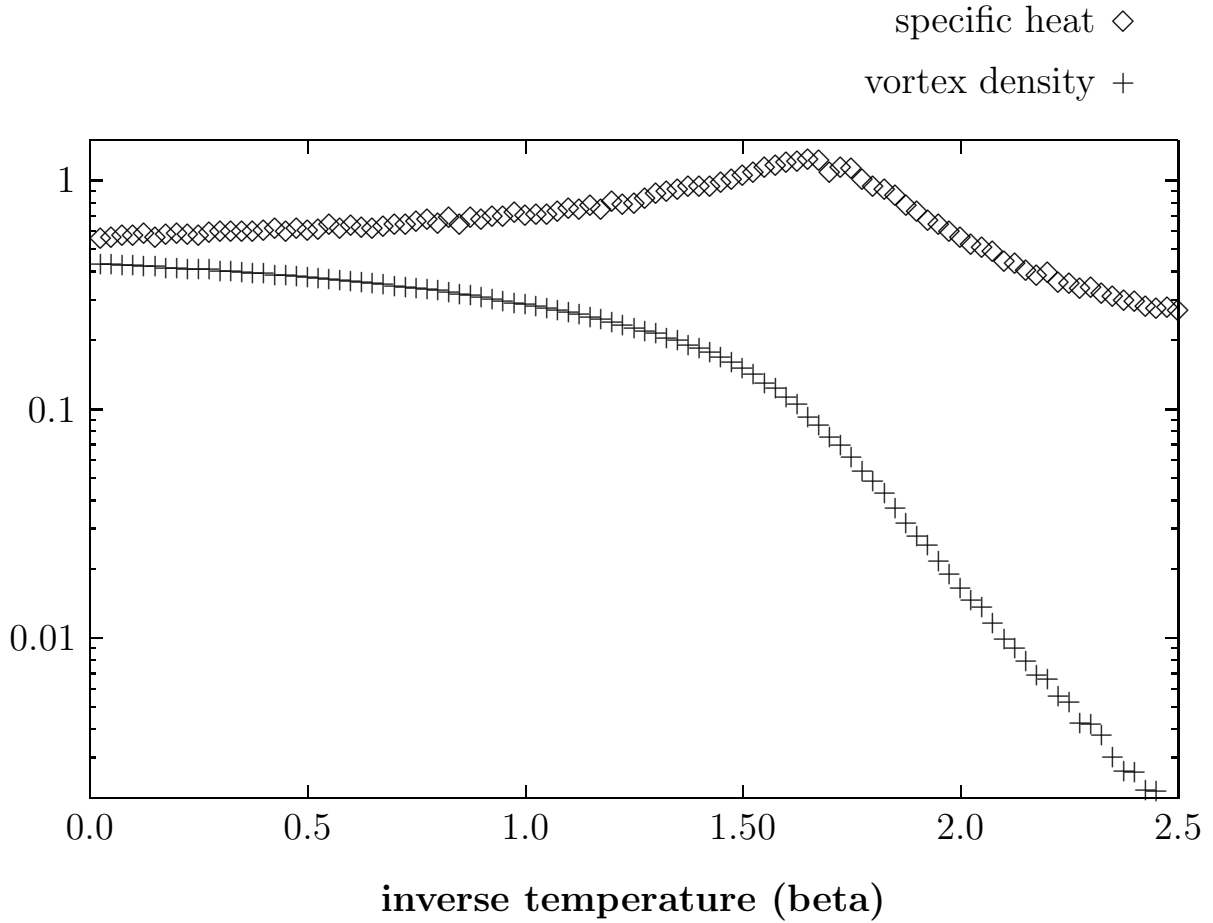
\begin{figure}[tb!]
\setlength{\unitlength}{0.240900pt}
\ifx\plotpoint\undefined\newsavebox{\plotpoint}\fi
\sbox{\plotpoint}{\rule[-0.200pt]{0.400pt}{0.400pt}}%
\begin{picture}(1949,1350)(100,0)
\font\gnuplot=cmr10 at 10pt
\gnuplot
\sbox{\plotpoint}{\rule[-0.200pt]{0.400pt}{0.400pt}}%
\put(176.0,113.0){\rule[-0.200pt]{0.400pt}{248.850pt}}
\put(176.0,113.0){\rule[-0.200pt]{2.409pt}{0.400pt}}
\put(1875.0,113.0){\rule[-0.200pt]{2.409pt}{0.400pt}}
\put(176.0,176.0){\rule[-0.200pt]{2.409pt}{0.400pt}}
\put(1875.0,176.0){\rule[-0.200pt]{2.409pt}{0.400pt}}
\put(176.0,221.0){\rule[-0.200pt]{2.409pt}{0.400pt}}
\put(1875.0,221.0){\rule[-0.200pt]{2.409pt}{0.400pt}}
\put(176.0,256.0){\rule[-0.200pt]{2.409pt}{0.400pt}}
\put(1875.0,256.0){\rule[-0.200pt]{2.409pt}{0.400pt}}
\put(176.0,284.0){\rule[-0.200pt]{2.409pt}{0.400pt}}
\put(1875.0,284.0){\rule[-0.200pt]{2.409pt}{0.400pt}}
\put(176.0,308.0){\rule[-0.200pt]{2.409pt}{0.400pt}}
\put(1875.0,308.0){\rule[-0.200pt]{2.409pt}{0.400pt}}
\put(176.0,329.0){\rule[-0.200pt]{2.409pt}{0.400pt}}
\put(1875.0,329.0){\rule[-0.200pt]{2.409pt}{0.400pt}}
\put(176.0,348.0){\rule[-0.200pt]{2.409pt}{0.400pt}}
\put(1875.0,348.0){\rule[-0.200pt]{2.409pt}{0.400pt}}
\put(176.0,364.0){\rule[-0.200pt]{4.818pt}{0.400pt}}
\put(154,364){\makebox(0,0)[r]{\large 0.01}}
\put(1865.0,364.0){\rule[-0.200pt]{4.818pt}{0.400pt}}
\put(176.0,472.0){\rule[-0.200pt]{2.409pt}{0.400pt}}
\put(1875.0,472.0){\rule[-0.200pt]{2.409pt}{0.400pt}}
\put(176.0,536.0){\rule[-0.200pt]{2.409pt}{0.400pt}}
\put(1875.0,536.0){\rule[-0.200pt]{2.409pt}{0.400pt}}
\put(176.0,580.0){\rule[-0.200pt]{2.409pt}{0.400pt}}
\put(1875.0,580.0){\rule[-0.200pt]{2.409pt}{0.400pt}}
\put(176.0,615.0){\rule[-0.200pt]{2.409pt}{0.400pt}}
\put(1875.0,615.0){\rule[-0.200pt]{2.409pt}{0.400pt}}
\put(176.0,644.0){\rule[-0.200pt]{2.409pt}{0.400pt}}
\put(1875.0,644.0){\rule[-0.200pt]{2.409pt}{0.400pt}}
\put(176.0,668.0){\rule[-0.200pt]{2.409pt}{0.400pt}}
\put(1875.0,668.0){\rule[-0.200pt]{2.409pt}{0.400pt}}
\put(176.0,689.0){\rule[-0.200pt]{2.409pt}{0.400pt}}
\put(1875.0,689.0){\rule[-0.200pt]{2.409pt}{0.400pt}}
\put(176.0,707.0){\rule[-0.200pt]{2.409pt}{0.400pt}}
\put(1875.0,707.0){\rule[-0.200pt]{2.409pt}{0.400pt}}
\put(176.0,723.0){\rule[-0.200pt]{4.818pt}{0.400pt}}
\put(154,723){\makebox(0,0)[r]{\large 0.1}}
\put(1865.0,723.0){\rule[-0.200pt]{4.818pt}{0.400pt}}
\put(176.0,832.0){\rule[-0.200pt]{2.409pt}{0.400pt}}
\put(1875.0,832.0){\rule[-0.200pt]{2.409pt}{0.400pt}}
\put(176.0,895.0){\rule[-0.200pt]{2.409pt}{0.400pt}}
\put(1875.0,895.0){\rule[-0.200pt]{2.409pt}{0.400pt}}
\put(176.0,940.0){\rule[-0.200pt]{2.409pt}{0.400pt}}
\put(1875.0,940.0){\rule[-0.200pt]{2.409pt}{0.400pt}}
\put(176.0,975.0){\rule[-0.200pt]{2.409pt}{0.400pt}}
\put(1875.0,975.0){\rule[-0.200pt]{2.409pt}{0.400pt}}
\put(176.0,1003.0){\rule[-0.200pt]{2.409pt}{0.400pt}}
\put(1875.0,1003.0){\rule[-0.200pt]{2.409pt}{0.400pt}}
\put(176.0,1027.0){\rule[-0.200pt]{2.409pt}{0.400pt}}
\put(1875.0,1027.0){\rule[-0.200pt]{2.409pt}{0.400pt}}
\put(176.0,1048.0){\rule[-0.200pt]{2.409pt}{0.400pt}}
\put(1875.0,1048.0){\rule[-0.200pt]{2.409pt}{0.400pt}}
\put(176.0,1066.0){\rule[-0.200pt]{2.409pt}{0.400pt}}
\put(1875.0,1066.0){\rule[-0.200pt]{2.409pt}{0.400pt}}
\put(176.0,1083.0){\rule[-0.200pt]{4.818pt}{0.400pt}}
\put(154,1083){\makebox(0,0)[r]{\large 1}}
\put(1865.0,1083.0){\rule[-0.200pt]{4.818pt}{0.400pt}}
\put(176.0,113.0){\rule[-0.200pt]{0.400pt}{4.818pt}}
\put(176,68){\makebox(0,0){\large 0.0}}
\put(176.0,1126.0){\rule[-0.200pt]{0.400pt}{4.818pt}}
\put(518.0,113.0){\rule[-0.200pt]{0.400pt}{4.818pt}}
\put(518,68){\makebox(0,0){\large 0.5}}
\put(518.0,1126.0){\rule[-0.200pt]{0.400pt}{4.818pt}}
\put(860.0,113.0){\rule[-0.200pt]{0.400pt}{4.818pt}}
\put(860,68){\makebox(0,0){\large 1.0}}
\put(860.0,1126.0){\rule[-0.200pt]{0.400pt}{4.818pt}}
\put(1201.0,113.0){\rule[-0.200pt]{0.400pt}{4.818pt}}
\put(1201,68){\makebox(0,0){\large 1.50}}
\put(1201.0,1126.0){\rule[-0.200pt]{0.400pt}{4.818pt}}
\put(1543.0,113.0){\rule[-0.200pt]{0.400pt}{4.818pt}}
\put(1543,68){\makebox(0,0){\large 2.0}}
\put(1543.0,1126.0){\rule[-0.200pt]{0.400pt}{4.818pt}}
\put(1885.0,113.0){\rule[-0.200pt]{0.400pt}{4.818pt}}
\put(1885,68){\makebox(0,0){\large 2.5}}
\put(1885.0,1126.0){\rule[-0.200pt]{0.400pt}{4.818pt}}
\put(176.0,113.0){\rule[-0.200pt]{411.698pt}{0.400pt}}
\put(1885.0,113.0){\rule[-0.200pt]{0.400pt}{248.850pt}}
\put(176.0,1146.0){\rule[-0.200pt]{411.698pt}{0.400pt}}
\put(1030,-50){\makebox(0,0){\bf \large inverse temperature (beta)}}
\put(176.0,113.0){\rule[-0.200pt]{0.400pt}{248.850pt}}
\put(1755,1331){\makebox(0,0)[r]{\large specific heat}}
\put(1799,1331){\raisebox{-.8pt}{\makebox(0,0){$\Diamond$}}}
\put(193,990){\raisebox{-.8pt}{\makebox(0,0){$\Diamond$}}}
\put(210,991){\raisebox{-.8pt}{\makebox(0,0){$\Diamond$}}}
\put(227,994){\raisebox{-.8pt}{\makebox(0,0){$\Diamond$}}}
\put(244,995){\raisebox{-.8pt}{\makebox(0,0){$\Diamond$}}}
\put(261,998){\raisebox{-.8pt}{\makebox(0,0){$\Diamond$}}}
\put(279,992){\raisebox{-.8pt}{\makebox(0,0){$\Diamond$}}}
\put(296,996){\raisebox{-.8pt}{\makebox(0,0){$\Diamond$}}}
\put(313,997){\raisebox{-.8pt}{\makebox(0,0){$\Diamond$}}}
\put(330,996){\raisebox{-.8pt}{\makebox(0,0){$\Diamond$}}}
\put(347,995){\raisebox{-.8pt}{\makebox(0,0){$\Diamond$}}}
\put(364,999){\raisebox{-.8pt}{\makebox(0,0){$\Diamond$}}}
\put(381,1000){\raisebox{-.8pt}{\makebox(0,0){$\Diamond$}}}
\put(398,1000){\raisebox{-.8pt}{\makebox(0,0){$\Diamond$}}}
\put(415,1000){\raisebox{-.8pt}{\makebox(0,0){$\Diamond$}}}
\put(432,1000){\raisebox{-.8pt}{\makebox(0,0){$\Diamond$}}}
\put(449,1003){\raisebox{-.8pt}{\makebox(0,0){$\Diamond$}}}
\put(467,1005){\raisebox{-.8pt}{\makebox(0,0){$\Diamond$}}}
\put(484,1001){\raisebox{-.8pt}{\makebox(0,0){$\Diamond$}}}
\put(501,1006){\raisebox{-.8pt}{\makebox(0,0){$\Diamond$}}}
\put(518,1002){\raisebox{-.8pt}{\makebox(0,0){$\Diamond$}}}
\put(535,1004){\raisebox{-.8pt}{\makebox(0,0){$\Diamond$}}}
\put(552,1011){\raisebox{-.8pt}{\makebox(0,0){$\Diamond$}}}
\put(569,1005){\raisebox{-.8pt}{\makebox(0,0){$\Diamond$}}}
\put(586,1010){\raisebox{-.8pt}{\makebox(0,0){$\Diamond$}}}
\put(603,1007){\raisebox{-.8pt}{\makebox(0,0){$\Diamond$}}}
\put(620,1006){\raisebox{-.8pt}{\makebox(0,0){$\Diamond$}}}
\put(637,1009){\raisebox{-.8pt}{\makebox(0,0){$\Diamond$}}}
\put(655,1012){\raisebox{-.8pt}{\makebox(0,0){$\Diamond$}}}
\put(672,1011){\raisebox{-.8pt}{\makebox(0,0){$\Diamond$}}}
\put(689,1017){\raisebox{-.8pt}{\makebox(0,0){$\Diamond$}}}
\put(706,1020){\raisebox{-.8pt}{\makebox(0,0){$\Diamond$}}}
\put(723,1014){\raisebox{-.8pt}{\makebox(0,0){$\Diamond$}}}
\put(740,1023){\raisebox{-.8pt}{\makebox(0,0){$\Diamond$}}}
\put(757,1012){\raisebox{-.8pt}{\makebox(0,0){$\Diamond$}}}
\put(774,1022){\raisebox{-.8pt}{\makebox(0,0){$\Diamond$}}}
\put(791,1019){\raisebox{-.8pt}{\makebox(0,0){$\Diamond$}}}
\put(808,1025){\raisebox{-.8pt}{\makebox(0,0){$\Diamond$}}}
\put(825,1024){\raisebox{-.8pt}{\makebox(0,0){$\Diamond$}}}
\put(843,1030){\raisebox{-.8pt}{\makebox(0,0){$\Diamond$}}}
\put(860,1026){\raisebox{-.8pt}{\makebox(0,0){$\Diamond$}}}
\put(877,1027){\raisebox{-.8pt}{\makebox(0,0){$\Diamond$}}}
\put(894,1027){\raisebox{-.8pt}{\makebox(0,0){$\Diamond$}}}
\put(911,1032){\raisebox{-.8pt}{\makebox(0,0){$\Diamond$}}}
\put(928,1037){\raisebox{-.8pt}{\makebox(0,0){$\Diamond$}}}
\put(945,1035){\raisebox{-.8pt}{\makebox(0,0){$\Diamond$}}}
\put(962,1042){\raisebox{-.8pt}{\makebox(0,0){$\Diamond$}}}
\put(979,1036){\raisebox{-.8pt}{\makebox(0,0){$\Diamond$}}}
\put(996,1048){\raisebox{-.8pt}{\makebox(0,0){$\Diamond$}}}
\put(1013,1043){\raisebox{-.8pt}{\makebox(0,0){$\Diamond$}}}
\put(1031,1045){\raisebox{-.8pt}{\makebox(0,0){$\Diamond$}}}
\put(1048,1053){\raisebox{-.8pt}{\makebox(0,0){$\Diamond$}}}
\put(1065,1061){\raisebox{-.8pt}{\makebox(0,0){$\Diamond$}}}
\put(1082,1063){\raisebox{-.8pt}{\makebox(0,0){$\Diamond$}}}
\put(1099,1066){\raisebox{-.8pt}{\makebox(0,0){$\Diamond$}}}
\put(1116,1071){\raisebox{-.8pt}{\makebox(0,0){$\Diamond$}}}
\put(1133,1071){\raisebox{-.8pt}{\makebox(0,0){$\Diamond$}}}
\put(1150,1072){\raisebox{-.8pt}{\makebox(0,0){$\Diamond$}}}
\put(1167,1078){\raisebox{-.8pt}{\makebox(0,0){$\Diamond$}}}
\put(1184,1083){\raisebox{-.8pt}{\makebox(0,0){$\Diamond$}}}
\put(1201,1089){\raisebox{-.8pt}{\makebox(0,0){$\Diamond$}}}
\put(1218,1093){\raisebox{-.8pt}{\makebox(0,0){$\Diamond$}}}
\put(1236,1101){\raisebox{-.8pt}{\makebox(0,0){$\Diamond$}}}
\put(1253,1104){\raisebox{-.8pt}{\makebox(0,0){$\Diamond$}}}
\put(1270,1109){\raisebox{-.8pt}{\makebox(0,0){$\Diamond$}}}
\put(1287,1110){\raisebox{-.8pt}{\makebox(0,0){$\Diamond$}}}
\put(1304,1114){\raisebox{-.8pt}{\makebox(0,0){$\Diamond$}}}
\put(1321,1112){\raisebox{-.8pt}{\makebox(0,0){$\Diamond$}}}
\put(1338,1093){\raisebox{-.8pt}{\makebox(0,0){$\Diamond$}}}
\put(1355,1101){\raisebox{-.8pt}{\makebox(0,0){$\Diamond$}}}
\put(1372,1099){\raisebox{-.8pt}{\makebox(0,0){$\Diamond$}}}
\put(1389,1083){\raisebox{-.8pt}{\makebox(0,0){$\Diamond$}}}
\put(1406,1072){\raisebox{-.8pt}{\makebox(0,0){$\Diamond$}}}
\put(1424,1066){\raisebox{-.8pt}{\makebox(0,0){$\Diamond$}}}
\put(1441,1057){\raisebox{-.8pt}{\makebox(0,0){$\Diamond$}}}
\put(1458,1042){\raisebox{-.8pt}{\makebox(0,0){$\Diamond$}}}
\put(1475,1032){\raisebox{-.8pt}{\makebox(0,0){$\Diamond$}}}
\put(1492,1018){\raisebox{-.8pt}{\makebox(0,0){$\Diamond$}}}
\put(1509,1011){\raisebox{-.8pt}{\makebox(0,0){$\Diamond$}}}
\put(1526,1001){\raisebox{-.8pt}{\makebox(0,0){$\Diamond$}}}
\put(1543,991){\raisebox{-.8pt}{\makebox(0,0){$\Diamond$}}}
\put(1560,981){\raisebox{-.8pt}{\makebox(0,0){$\Diamond$}}}
\put(1577,976){\raisebox{-.8pt}{\makebox(0,0){$\Diamond$}}}
\put(1594,969){\raisebox{-.8pt}{\makebox(0,0){$\Diamond$}}}
\put(1612,954){\raisebox{-.8pt}{\makebox(0,0){$\Diamond$}}}
\put(1629,950){\raisebox{-.8pt}{\makebox(0,0){$\Diamond$}}}
\put(1646,940){\raisebox{-.8pt}{\makebox(0,0){$\Diamond$}}}
\put(1663,931){\raisebox{-.8pt}{\makebox(0,0){$\Diamond$}}}
\put(1680,936){\raisebox{-.8pt}{\makebox(0,0){$\Diamond$}}}
\put(1697,921){\raisebox{-.8pt}{\makebox(0,0){$\Diamond$}}}
\put(1714,919){\raisebox{-.8pt}{\makebox(0,0){$\Diamond$}}}
\put(1731,911){\raisebox{-.8pt}{\makebox(0,0){$\Diamond$}}}
\put(1748,913){\raisebox{-.8pt}{\makebox(0,0){$\Diamond$}}}
\put(1765,904){\raisebox{-.8pt}{\makebox(0,0){$\Diamond$}}}
\put(1782,898){\raisebox{-.8pt}{\makebox(0,0){$\Diamond$}}}
\put(1800,893){\raisebox{-.8pt}{\makebox(0,0){$\Diamond$}}}
\put(1817,891){\raisebox{-.8pt}{\makebox(0,0){$\Diamond$}}}
\put(1834,883){\raisebox{-.8pt}{\makebox(0,0){$\Diamond$}}}
\put(1851,880){\raisebox{-.8pt}{\makebox(0,0){$\Diamond$}}}
\put(1868,882){\raisebox{-.8pt}{\makebox(0,0){$\Diamond$}}}
\put(1885,877){\raisebox{-.8pt}{\makebox(0,0){$\Diamond$}}}
\put(1755,1236){\makebox(0,0)[r]{\large vortex density}}
\put(1799,1236){\makebox(0,0){$+$}}
\put(193,951){\makebox(0,0){$+$}}
\put(210,951){\makebox(0,0){$+$}}
\put(227,950){\makebox(0,0){$+$}}
\put(244,950){\makebox(0,0){$+$}}
\put(261,948){\makebox(0,0){$+$}}
\put(279,948){\makebox(0,0){$+$}}
\put(296,946){\makebox(0,0){$+$}}
\put(313,946){\makebox(0,0){$+$}}
\put(330,944){\makebox(0,0){$+$}}
\put(347,944){\makebox(0,0){$+$}}
\put(364,943){\makebox(0,0){$+$}}
\put(381,941){\makebox(0,0){$+$}}
\put(398,940){\makebox(0,0){$+$}}
\put(415,939){\makebox(0,0){$+$}}
\put(432,938){\makebox(0,0){$+$}}
\put(449,937){\makebox(0,0){$+$}}
\put(467,935){\makebox(0,0){$+$}}
\put(484,934){\makebox(0,0){$+$}}
\put(501,932){\makebox(0,0){$+$}}
\put(518,931){\makebox(0,0){$+$}}
\put(535,930){\makebox(0,0){$+$}}
\put(552,928){\makebox(0,0){$+$}}
\put(569,926){\makebox(0,0){$+$}}
\put(586,925){\makebox(0,0){$+$}}
\put(603,923){\makebox(0,0){$+$}}
\put(620,921){\makebox(0,0){$+$}}
\put(637,920){\makebox(0,0){$+$}}
\put(655,917){\makebox(0,0){$+$}}
\put(672,916){\makebox(0,0){$+$}}
\put(689,914){\makebox(0,0){$+$}}
\put(706,912){\makebox(0,0){$+$}}
\put(723,910){\makebox(0,0){$+$}}
\put(740,907){\makebox(0,0){$+$}}
\put(757,905){\makebox(0,0){$+$}}
\put(774,903){\makebox(0,0){$+$}}
\put(791,900){\makebox(0,0){$+$}}
\put(808,897){\makebox(0,0){$+$}}
\put(825,894){\makebox(0,0){$+$}}
\put(843,891){\makebox(0,0){$+$}}
\put(860,889){\makebox(0,0){$+$}}
\put(877,886){\makebox(0,0){$+$}}
\put(894,883){\makebox(0,0){$+$}}
\put(911,879){\makebox(0,0){$+$}}
\put(928,876){\makebox(0,0){$+$}}
\put(945,873){\makebox(0,0){$+$}}
\put(962,868){\makebox(0,0){$+$}}
\put(979,865){\makebox(0,0){$+$}}
\put(996,860){\makebox(0,0){$+$}}
\put(1013,856){\makebox(0,0){$+$}}
\put(1031,851){\makebox(0,0){$+$}}
\put(1048,847){\makebox(0,0){$+$}}
\put(1065,843){\makebox(0,0){$+$}}
\put(1082,836){\makebox(0,0){$+$}}
\put(1099,832){\makebox(0,0){$+$}}
\put(1116,825){\makebox(0,0){$+$}}
\put(1133,820){\makebox(0,0){$+$}}
\put(1150,813){\makebox(0,0){$+$}}
\put(1167,805){\makebox(0,0){$+$}}
\put(1184,798){\makebox(0,0){$+$}}
\put(1201,788){\makebox(0,0){$+$}}
\put(1218,778){\makebox(0,0){$+$}}
\put(1236,765){\makebox(0,0){$+$}}
\put(1253,756){\makebox(0,0){$+$}}
\put(1270,743){\makebox(0,0){$+$}}
\put(1287,732){\makebox(0,0){$+$}}
\put(1304,711){\makebox(0,0){$+$}}
\put(1321,698){\makebox(0,0){$+$}}
\put(1338,680){\makebox(0,0){$+$}}
\put(1355,667){\makebox(0,0){$+$}}
\put(1372,649){\makebox(0,0){$+$}}
\put(1389,626){\makebox(0,0){$+$}}
\put(1406,611){\makebox(0,0){$+$}}
\put(1424,591){\makebox(0,0){$+$}}
\put(1441,568){\makebox(0,0){$+$}}
\put(1458,544){\makebox(0,0){$+$}}
\put(1475,524){\makebox(0,0){$+$}}
\put(1492,510){\makebox(0,0){$+$}}
\put(1509,485){\makebox(0,0){$+$}}
\put(1526,464){\makebox(0,0){$+$}}
\put(1543,443){\makebox(0,0){$+$}}
\put(1560,423){\makebox(0,0){$+$}}
\put(1577,412){\makebox(0,0){$+$}}
\put(1594,388){\makebox(0,0){$+$}}
\put(1612,362){\makebox(0,0){$+$}}
\put(1629,348){\makebox(0,0){$+$}}
\put(1646,328){\makebox(0,0){$+$}}
\put(1663,306){\makebox(0,0){$+$}}
\put(1680,300){\makebox(0,0){$+$}}
\put(1697,273){\makebox(0,0){$+$}}
\put(1714,264){\makebox(0,0){$+$}}
\put(1731,231){\makebox(0,0){$+$}}
\put(1748,229){\makebox(0,0){$+$}}
\put(1765,212){\makebox(0,0){$+$}}
\put(1782,177){\makebox(0,0){$+$}}
\put(1800,155){\makebox(0,0){$+$}}
\put(1817,153){\makebox(0,0){$+$}}
\put(1834,125){\makebox(0,0){$+$}}
\put(1851,123){\makebox(0,0){$+$}}
\end{picture}
\vspace{2mm}
\caption{\small The specific heat and the 
vortex density versus inverse temperature
($\beta$) on an 8$\times$8 lattice calculated with the SO(3) action.
 \label{fig:specheat}}
\end{figure}

It is well-known that both in 
non-Abelian 2D spin models and the analogous 4D gauge theories the strong
and weak coupling regions are separated by a crossover region, where the
specific heat has a finite peak \cite{crossover}. We used the specific heat
peak to ensure that the coupling that we use in the simulations is already
``weak''. The Monte Carlo simulation was performed at $\beta=2.0$ which is
already in the weak coupling region as can be seen from Figure
\ref{fig:specheat}. We also plotted in the same Figure the Z(2) vortex 
density versus the inverse temperature. Another characteristic feature
of the weak coupling region is that the vortex density (monopole density
in gauge theories) decreases exponentially as a function of $\beta$ 
\cite{crossover}. This can also be clearly seen in the Figure.

We do not expect any dramatic phenomenon to happen 
when going from the weak coupling side of the specific heat peak to even
weaker and weaker couplings, therefore our results should qualitatively apply 
for any ``weak'' but of course non-zero coupling.

We measured the probability of the configuration with two adjecent vortices
at fixed positions separated by $C_2$, normalised by the probability of 
having no vortices at all. The calculation was done by generating a series
of configurations using a local heat bath algorithm with the SO(3) invariant
measure $d\nu[U]$ but {\em without the constraint on the $\eta$ string}.
In a long Monte Carlo run the number of configurations of the type in Figure
\ref{fig:gaploopa} was counted and divided by the number of configurations 
with no vortices at all and an even number of $\eta$ strings crossing $C_2$.

Since (at fixed $\beta$) on larger lattices there are typically more vortices,
the probability of the configurations with exactly zero and one vortex pair
decreased very rapidly with increasing lattice size. This meant that the 
quantity we wanted to measure was given as a ratio between two numbers both
becoming very small on larger lattices, however their ratio was expected
to be stable. This made the signal less accurate on larger lattices.
An improvement
by a factor $L^2$ could be achieved by counting all the configurations 
that were translations of the one in Figure \ref{fig:gaploopa} and 
dividing the result by $L^2$. Because of the periodic boundary conditions
the lattice had an exact translation invariance and this procedure 
did not change the results.

\begin{figure}[tb!]
\setlength{\unitlength}{0.240900pt}
\ifx\plotpoint\undefined\newsavebox{\plotpoint}\fi
\sbox{\plotpoint}{\rule[-0.200pt]{0.400pt}{0.400pt}}%
\begin{picture}(1949,1169)(100,200)
\font\gnuplot=cmr10 at 10pt
\gnuplot
\sbox{\plotpoint}{\rule[-0.200pt]{0.400pt}{0.400pt}}%
\put(176.0,241.0){\rule[-0.200pt]{4.818pt}{0.400pt}}
\put(154,241){\makebox(0,0)[r]{\large 2e-07}}
\put(1865.0,241.0){\rule[-0.200pt]{4.818pt}{0.400pt}}
\put(176.0,392.0){\rule[-0.200pt]{4.818pt}{0.400pt}}
\put(154,392){\makebox(0,0)[r]{\large 4e-07}}
\put(1865.0,392.0){\rule[-0.200pt]{4.818pt}{0.400pt}}
\put(176.0,543.0){\rule[-0.200pt]{4.818pt}{0.400pt}}
\put(154,543){\makebox(0,0)[r]{\large 6e-07}}
\put(1865.0,543.0){\rule[-0.200pt]{4.818pt}{0.400pt}}
\put(176.0,694.0){\rule[-0.200pt]{4.818pt}{0.400pt}}
\put(154,694){\makebox(0,0)[r]{\large 8e-07}}
\put(1865.0,694.0){\rule[-0.200pt]{4.818pt}{0.400pt}}
\put(176.0,844.0){\rule[-0.200pt]{4.818pt}{0.400pt}}
\put(154,844){\makebox(0,0)[r]{\large 1e-06}}
\put(1865.0,844.0){\rule[-0.200pt]{4.818pt}{0.400pt}}
\put(176.0,995.0){\rule[-0.200pt]{4.818pt}{0.400pt}}
\put(154,995){\makebox(0,0)[r]{\large 1.2e-06}}
\put(1865.0,995.0){\rule[-0.200pt]{4.818pt}{0.400pt}}
\put(176.0,1146.0){\rule[-0.200pt]{4.818pt}{0.400pt}}
\put(154,1146){\makebox(0,0)[r]{\large 1.4e-06}}
\put(1865.0,1146.0){\rule[-0.200pt]{4.818pt}{0.400pt}}
\put(176.0,113.0){\rule[-0.200pt]{0.400pt}{4.818pt}}
\put(176,68){\makebox(0,0){\large 4}}
\put(176.0,1126.0){\rule[-0.200pt]{0.400pt}{4.818pt}}
\put(518.0,113.0){\rule[-0.200pt]{0.400pt}{4.818pt}}
\put(518,68){\makebox(0,0){\large 6}}
\put(518.0,1126.0){\rule[-0.200pt]{0.400pt}{4.818pt}}
\put(860.0,113.0){\rule[-0.200pt]{0.400pt}{4.818pt}}
\put(860,68){\makebox(0,0){\large 8}}
\put(860.0,1126.0){\rule[-0.200pt]{0.400pt}{4.818pt}}
\put(1201.0,113.0){\rule[-0.200pt]{0.400pt}{4.818pt}}
\put(1201,68){\makebox(0,0){\large 10}}
\put(1201.0,1126.0){\rule[-0.200pt]{0.400pt}{4.818pt}}
\put(1543.0,113.0){\rule[-0.200pt]{0.400pt}{4.818pt}}
\put(1543,68){\makebox(0,0){\large 12}}
\put(1543.0,1126.0){\rule[-0.200pt]{0.400pt}{4.818pt}}
\put(1885.0,113.0){\rule[-0.200pt]{0.400pt}{4.818pt}}
\put(1885,68){\makebox(0,0){\large 14}}
\put(1885.0,1126.0){\rule[-0.200pt]{0.400pt}{4.818pt}}
\put(176.0,113.0){\rule[-0.200pt]{411.698pt}{0.400pt}}
\put(1885.0,113.0){\rule[-0.200pt]{0.400pt}{248.850pt}}
\put(176.0,1146.0){\rule[-0.200pt]{411.698pt}{0.400pt}}
\put(1030,23){\makebox(0,-100){\Large lattice size}}
\put(176.0,113.0){\rule[-0.200pt]{0.400pt}{248.850pt}}
\put(347,942){\raisebox{-.8pt}{\makebox(0,0){$\Diamond$}}}
\put(518,492){\raisebox{-.8pt}{\makebox(0,0){$\Diamond$}}}
\put(689,384){\raisebox{-.8pt}{\makebox(0,0){$\Diamond$}}}
\put(860,308){\raisebox{-.8pt}{\makebox(0,0){$\Diamond$}}}
\put(1031,258){\raisebox{-.8pt}{\makebox(0,0){$\Diamond$}}}
\put(1201,269){\raisebox{-.8pt}{\makebox(0,0){$\Diamond$}}}
\put(1372,252){\raisebox{-.8pt}{\makebox(0,0){$\Diamond$}}}
\put(1543,265){\raisebox{-.8pt}{\makebox(0,0){$\Diamond$}}}
\put(1714,263){\raisebox{-.8pt}{\makebox(0,0){$\Diamond$}}}
\put(347.0,867.0){\rule[-0.200pt]{0.400pt}{36.376pt}}
\put(337.0,867.0){\rule[-0.200pt]{4.818pt}{0.400pt}}
\put(337.0,1018.0){\rule[-0.200pt]{4.818pt}{0.400pt}}
\put(518.0,457.0){\rule[-0.200pt]{0.400pt}{17.104pt}}
\put(508.0,457.0){\rule[-0.200pt]{4.818pt}{0.400pt}}
\put(508.0,528.0){\rule[-0.200pt]{4.818pt}{0.400pt}}
\put(689.0,327.0){\rule[-0.200pt]{0.400pt}{27.222pt}}
\put(679.0,327.0){\rule[-0.200pt]{4.818pt}{0.400pt}}
\put(679.0,440.0){\rule[-0.200pt]{4.818pt}{0.400pt}}
\put(860.0,268.0){\rule[-0.200pt]{0.400pt}{19.031pt}}
\put(850.0,268.0){\rule[-0.200pt]{4.818pt}{0.400pt}}
\put(850.0,347.0){\rule[-0.200pt]{4.818pt}{0.400pt}}
\put(1031.0,219.0){\rule[-0.200pt]{0.400pt}{18.790pt}}
\put(1021.0,219.0){\rule[-0.200pt]{4.818pt}{0.400pt}}
\put(1021.0,297.0){\rule[-0.200pt]{4.818pt}{0.400pt}}
\put(1201.0,251.0){\rule[-0.200pt]{0.400pt}{8.672pt}}
\put(1191.0,251.0){\rule[-0.200pt]{4.818pt}{0.400pt}}
\put(1191.0,287.0){\rule[-0.200pt]{4.818pt}{0.400pt}}
\put(1372.0,231.0){\rule[-0.200pt]{0.400pt}{9.877pt}}
\put(1362.0,231.0){\rule[-0.200pt]{4.818pt}{0.400pt}}
\put(1362.0,272.0){\rule[-0.200pt]{4.818pt}{0.400pt}}
\put(1543.0,237.0){\rule[-0.200pt]{0.400pt}{13.731pt}}
\put(1533.0,237.0){\rule[-0.200pt]{4.818pt}{0.400pt}}
\put(1533.0,294.0){\rule[-0.200pt]{4.818pt}{0.400pt}}
\put(1714.0,235.0){\rule[-0.200pt]{0.400pt}{13.490pt}}
\put(1704.0,235.0){\rule[-0.200pt]{4.818pt}{0.400pt}}
\put(1704.0,291.0){\rule[-0.200pt]{4.818pt}{0.400pt}}
\end{picture}
\vspace{12mm}
\caption{\small The probability of a neighbouring vortex pair with 
its $\eta$ string winding around the lattice (see Fig \ref{fig:gaploopa})
as a function of the lattice size at $\beta=2.0$. The measurement was
done with the SO(3) action.
\label{fig:Fpair}}
\end{figure}
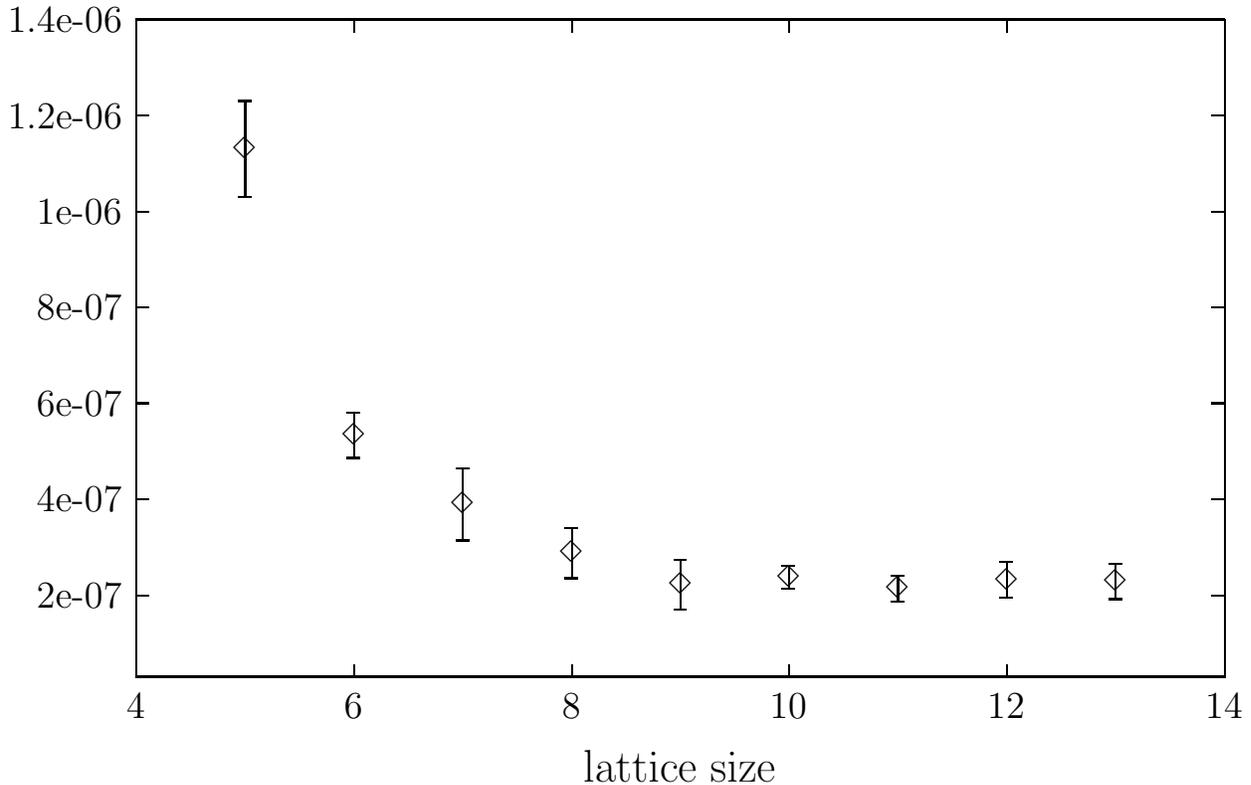

The simulations were performed on square lattices of $5 \leq L \leq 13$.
The rapidly deteriorating quality of the 
signal made it impossible to go beyond $L=13$. Even at
this point we typically needed several hundred thousands of independent 
configurations to get a signal at all. Our results are summarised in Figure
\ref{fig:Fpair}. We can see that the probability of a vortex pair with a 
long $\eta$ string decreases on small lattices until it stabilises at a 
moderate lattice size $(L \approx 8-9)$ and stays constant thereafter. 
Recall that for our purposes it is enough if this quantity remains non-zero
in the $L \rightarrow \infty$ limit.

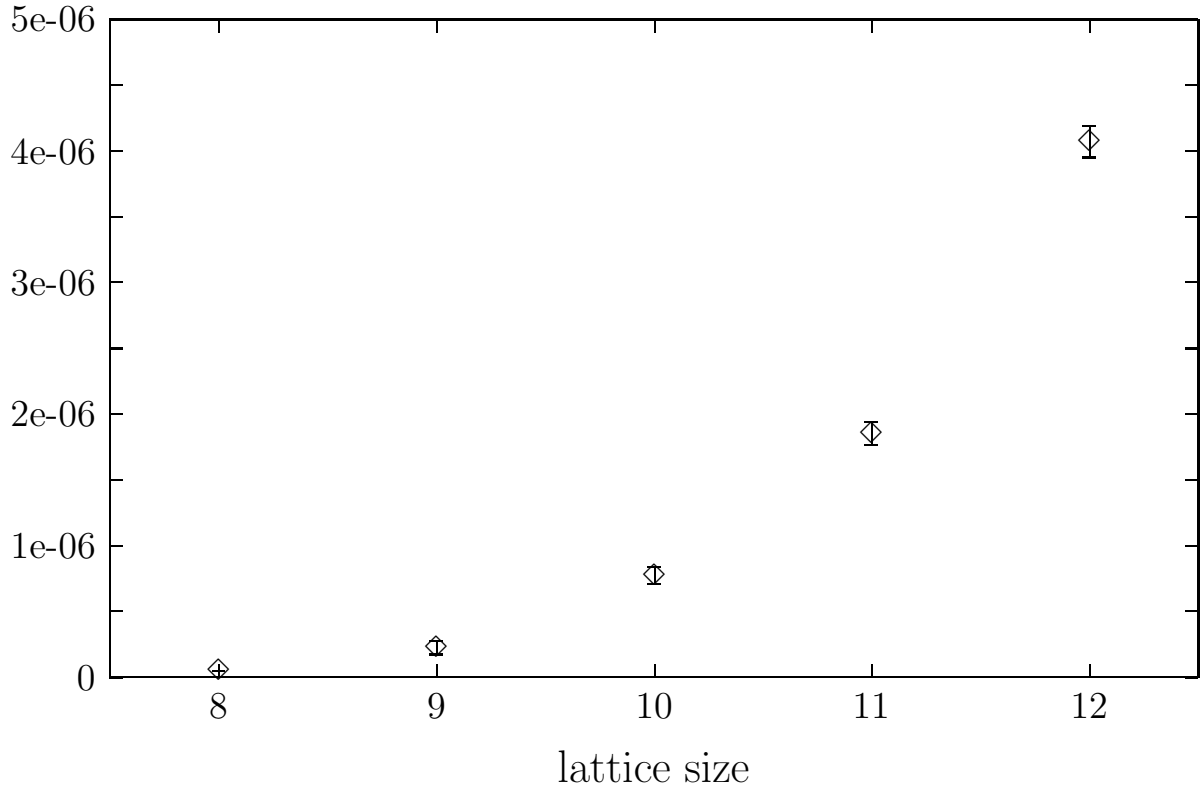
\begin{figure}[tb!]
\setlength{\unitlength}{0.240900pt}
\ifx\plotpoint\undefined\newsavebox{\plotpoint}\fi
\sbox{\plotpoint}{\rule[-0.200pt]{0.400pt}{0.400pt}}%
\begin{picture}(1949,1169)(100,200)
\font\gnuplot=cmr10 at 10pt
\gnuplot
\sbox{\plotpoint}{\rule[-0.200pt]{0.400pt}{0.400pt}}%
\put(176.0,112.0){\rule[-0.200pt]{4.818pt}{0.400pt}}
\put(154,112){\makebox(0,0)[r]{\large 0}}
\put(1865.0,112.0){\rule[-0.200pt]{4.818pt}{0.400pt}}
\put(176.0,216.0){\rule[-0.200pt]{4.818pt}{0.400pt}}
\put(1865.0,216.0){\rule[-0.200pt]{4.818pt}{0.400pt}}
\put(176.0,319.0){\rule[-0.200pt]{4.818pt}{0.400pt}}
\put(154,319){\makebox(0,0)[r]{\large 1e-06}}
\put(1865.0,319.0){\rule[-0.200pt]{4.818pt}{0.400pt}}
\put(176.0,422.0){\rule[-0.200pt]{4.818pt}{0.400pt}}
\put(1865.0,422.0){\rule[-0.200pt]{4.818pt}{0.400pt}}
\put(176.0,526.0){\rule[-0.200pt]{4.818pt}{0.400pt}}
\put(154,526){\makebox(0,0)[r]{\large 2e-06}}
\put(1865.0,526.0){\rule[-0.200pt]{4.818pt}{0.400pt}}
\put(176.0,629.0){\rule[-0.200pt]{4.818pt}{0.400pt}}
\put(1865.0,629.0){\rule[-0.200pt]{4.818pt}{0.400pt}}
\put(176.0,733.0){\rule[-0.200pt]{4.818pt}{0.400pt}}
\put(154,733){\makebox(0,0)[r]{\large 3e-06}}
\put(1865.0,733.0){\rule[-0.200pt]{4.818pt}{0.400pt}}
\put(176.0,836.0){\rule[-0.200pt]{4.818pt}{0.400pt}}
\put(1865.0,836.0){\rule[-0.200pt]{4.818pt}{0.400pt}}
\put(176.0,939.0){\rule[-0.200pt]{4.818pt}{0.400pt}}
\put(154,939){\makebox(0,0)[r]{\large 4e-06}}
\put(1865.0,939.0){\rule[-0.200pt]{4.818pt}{0.400pt}}
\put(176.0,1043.0){\rule[-0.200pt]{4.818pt}{0.400pt}}
\put(1865.0,1043.0){\rule[-0.200pt]{4.818pt}{0.400pt}}
\put(176.0,1146.0){\rule[-0.200pt]{4.818pt}{0.400pt}}
\put(154,1146){\makebox(0,0)[r]{\large 5e-06}}
\put(1865.0,1146.0){\rule[-0.200pt]{4.818pt}{0.400pt}}
\put(347.0,113.0){\rule[-0.200pt]{0.400pt}{4.818pt}}
\put(347,68){\makebox(0,0){\large 8}}
\put(347.0,1126.0){\rule[-0.200pt]{0.400pt}{4.818pt}}
\put(689.0,113.0){\rule[-0.200pt]{0.400pt}{4.818pt}}
\put(689,68){\makebox(0,0){\large 9}}
\put(689.0,1126.0){\rule[-0.200pt]{0.400pt}{4.818pt}}
\put(1031.0,113.0){\rule[-0.200pt]{0.400pt}{4.818pt}}
\put(1031,68){\makebox(0,0){\large 10}}
\put(1031.0,1126.0){\rule[-0.200pt]{0.400pt}{4.818pt}}
\put(1372.0,113.0){\rule[-0.200pt]{0.400pt}{4.818pt}}
\put(1372,68){\makebox(0,0){\large 11}}
\put(1372.0,1126.0){\rule[-0.200pt]{0.400pt}{4.818pt}}
\put(1714.0,113.0){\rule[-0.200pt]{0.400pt}{4.818pt}}
\put(1714,68){\makebox(0,0){\large 12}}
\put(1714.0,1126.0){\rule[-0.200pt]{0.400pt}{4.818pt}}
\put(176.0,113.0){\rule[-0.200pt]{411.698pt}{0.400pt}}
\put(1885.0,113.0){\rule[-0.200pt]{0.400pt}{248.850pt}}
\put(176.0,1146.0){\rule[-0.200pt]{411.698pt}{0.400pt}}
\put(1030,23){\makebox(0,-100){\Large lattice size}}
\put(176.0,113.0){\rule[-0.200pt]{0.400pt}{248.850pt}}
\put(347,122){\raisebox{-.8pt}{\makebox(0,0){$\Diamond$}}}
\put(689,158){\raisebox{-.8pt}{\makebox(0,0){$\Diamond$}}}
\put(1031,272){\raisebox{-.8pt}{\makebox(0,0){$\Diamond$}}}
\put(1372,495){\raisebox{-.8pt}{\makebox(0,0){$\Diamond$}}}
\put(1714,954){\raisebox{-.8pt}{\makebox(0,0){$\Diamond$}}}
\put(347,122){\usebox{\plotpoint}}
\put(337.0,122.0){\rule[-0.200pt]{4.818pt}{0.400pt}}
\put(337.0,122.0){\rule[-0.200pt]{4.818pt}{0.400pt}}
\put(689.0,148.0){\rule[-0.200pt]{0.400pt}{5.059pt}}
\put(679.0,148.0){\rule[-0.200pt]{4.818pt}{0.400pt}}
\put(679.0,169.0){\rule[-0.200pt]{4.818pt}{0.400pt}}
\put(1031.0,259.0){\rule[-0.200pt]{0.400pt}{6.504pt}}
\put(1021.0,259.0){\rule[-0.200pt]{4.818pt}{0.400pt}}
\put(1021.0,286.0){\rule[-0.200pt]{4.818pt}{0.400pt}}
\put(1372.0,477.0){\rule[-0.200pt]{0.400pt}{8.672pt}}
\put(1362.0,477.0){\rule[-0.200pt]{4.818pt}{0.400pt}}
\put(1362.0,513.0){\rule[-0.200pt]{4.818pt}{0.400pt}}
\put(1714.0,929.0){\rule[-0.200pt]{0.400pt}{12.045pt}}
\put(1704.0,929.0){\rule[-0.200pt]{4.818pt}{0.400pt}}
\put(1704.0,979.0){\rule[-0.200pt]{4.818pt}{0.400pt}}
\end{picture}
\vspace{1cm}
\caption{\small The same as in Figure \ref{fig:Fpair} except that instead of
using a square lattice the length in the direction parallel to the
$\eta$ string was fixed to 9 and only the transverse (perpendicular to the
string) size was varied.
\label{fig:spread1}}
\end{figure}

To study the spreading of the flux of the long eta string, we measured 
the same quantity with the lattice size fixed in the direction parallel 
to the string and varied in the perpendicular direction. The results are 
plotted in Figure \ref{fig:spread1}. The pronounced effect of flux spreding 
is obvious; as the lattice becomes wider, there is more space available
for the flux of the $\eta$ string to spread and its probability increases
dramatically.

To summarise, our Monte Carlo results are qualitatively consistent with the
semiclassical picture of flux spreading and it is indeed very hard to imagine
that anything could happen either at higher $\beta$ or larger lattice sizes
that could make the two-vortex probability vanish in the $L \rightarrow
\infty$ limit. Of course our results do not imply any particular analytic
form as to how the flux actually spreads and it seems very hard to distinguish
between a semiclassical ``massless'' spreading and an exponential spreading.
This is however not necessary for our purposes; the only property that 
we need is that the probability of the long string with two vortices 
remain non-zero in the $L \rightarrow \infty$ limit.

\section{Conclusions}

We proposed a mechanism that is sufficient to create a nonzero mass gap in 
the SU(2) chiral spin model at arbitrarily small temperatures. Instead of 
considering the correlation function we were looking at an operator ($G(L)$,
the twist) that measured how effective fluctuations were in destroying the
correlations between the ``relative signs'' of spins at different locations.

In this way we could separate the spin system
into two interacting parts, a Z(2) and an SO(3). We derived the conditions
that the SO(3) part had to satisfy so that the mass gap could be rigorously
established in terms of an effective Z(2) system, which was an
Ising model on the dual lattice. By using a plausible (but so far not proved)
vortex correlation inequality we could reduce these necessary conditions to
just one condition, namely the boundedness of the free energy of a vortex
pair connected by an $\eta$ string winding around the lattice for any lattice
size.

This condition on the truly non-Abelian part of the system was
substantially weaker than the existence of a mass gap itself. The presence
of a mass gap would require that an external Z(2) flux, introduced by twisted
boundary conditions, spread exponentially fast on large lattices, i.e.
the expectation of a twist would have to go to 1 exponentially. On the other
hand in our scheme we only needed that the expectation of the twist be 
non-zero for asymptotically large lattices. It is nevertheless surprising
that we failed to verify even this substantially weaker condition rigorously
and eventually we had to resort to Monte Carlo to check it. As expected,
our Monte Carlo data show that even at low temperature the expectation of
the twist goes to a nonzero constant as $L \rightarrow \infty$. It would
be however very desirable to have an analytic proof of this and the vortex
correlation inequalities too. 

This framework can also be applied to
3d and 4d SU(2) gauge theories with essentially no modifications. The only
minor difference here is that the corresponding objects on the lattice live
on higher dimensional simplices. For example instead of Z(2) vortices
on plaquettes we have Z(2) monopoles (monopole loops in 4d) on cubes and
the twist in the spin model is exactly analogous to the sourceless 't Hooft
loop in gauge theories (for a definition see \cite{Tomb-Yaffe}). 
In this way the confinement problem can be reduced 
to Z(2) monopole correlation inequalities. Monte Carlo results concerning 
the relevant monopole correlations will be reported elsewhere \cite{prep}.
This gives a unifying picture of confinement in gauge 
theories and disorder in spin models.

To summarise, we rigorously proved that the presence of a mass gap in the
2d SU(2) chiral spin model at low temperature is tantamount to certain
vortex correlation inequalities in a related SO(3) spin model. These
inequalities in turn have been (partly) checked by Monte Carlo calculations.
It would be worthwile to obtain an analytic proof of the vortex correlation 
inequalities and thus complete the solution of this 
longstanding problem. Finally, it
would be also interesting to extend this scheme to other symmetry groups,
in particular to SU(n).

\section*{Acknowledgement}

I am grateful to Terry Tomboulis for introducing me to this problem and
also for many helpful conversations. I also thank the Department of 
Theoretical Physics, Kossuth University, Debrecen, Hungary for granting
me part of the computer time used for the Monte Carlo.

\section*{Appendix}

 In the Appendix we derive the decomposition of $\Zeff Z(\Geff -G)$
into a sum of vortex expectations with non-negative coefficients.
Using (\ref{eq:Geff}) and (\ref{eq:Gomega}) (with the measure $d\nu_+[U]$
and $K(\mbox{d}U_l)$ computed from (\ref{eq:KMnew})) 
this can be written as a double Z(2) integral for two independent copies of
the $\omega$ variables on each plaquette,
\begin{eqnarray}
 Z \, \Zeff \, (\Geff - G)  =  \int\!\!d\nu_+[U] \, \prod_{p \in \Lambda}
 \int\!\!d\omega_p \int\!\!d\omtld_p \, \chi_{\omega_p}(\mbox{d}\eta_p) \,
 e^{h \omtld_p}  \; \exp \sum_{l \notin C_2} K \left( \dstar 
 \omega_l + \dstar \omtld_l \right)  
          \nonumber \\
 \times \left[ \exp \sum_{l \in C_2} K \left( \dstar \omega_l - 
 \dstar \omtld_l \right) \; -  \; \exp -\sum_{l \in C_2} K \left( \dstar 
 \omega_l - \dstar \omtld_l \right) \right].
     \label{eq:GmG}
\end{eqnarray}
We shall make use of the identity
\begin{equation}
 \prod_{i=1}^n f_i - \prod_{i=1}^n g_i = \frac{1}{2^{n-1}}
 \sum_{\{+-...\}} \prod_{i=1}^n  \left( f_i \pm g_i \right),
\end{equation}
where the summation is on all possible strings of plus and minus signs of
length $n$ containing an odd number of minus signs. This can be easily verified
by induction on $n$ (see \cite{Ginibre}). 

Now by applying this identity on the expression in the second line of
(\ref{eq:GmG}) it can be written as
\begin{eqnarray}
 \lefteqn{\exp K \sum_{l \in C_2} \left( \dstar \omega_l - 
 \dstar \omtld_l \right) \; -  \; \exp -K \sum_{l \in C_2} \left( 
 \dstar \omega_l - \dstar \omtld_l \right) = } \hspace{4cm}
                      \nonumber \\
 & & \frac{1}{2^{L-1}} \sum_{\{+-..\} }
 \prod_{l \in C_2} \left[ e^{K(\sd^\star \omega_l -\sd^\star \omtld_l )}
 \pm e^{-K(\sd^\star \omega_l -\sd^\star \omtld_l} \right].
     \label{eq:C2links}
\end{eqnarray}
We can now expand all the factors in the integrand of (\ref{eq:GmG}) using
\begin{equation}
 f(\omega) = \hat{f}(1) \left( 1 + \omega \frac{\hat{f}(-1)}{\hat{f}(1)}
 \right),
\end{equation}
(see equation (\ref{eq:chexp})) and then make a change of variables 
\begin{equation}
 \omp = \frac{1}{2} (\omega + \omtld) \hspace{2cm} \mbox{and} \hspace{2cm}
 \omm = \frac{1}{2} (\omega - \omtld)
     \label{eq:chvar}
\end{equation}
on each plaquette. This will result in an expression containing factors
of the form $(\omp)^{n_p} (\omm)^{m_p}$ on each plaquette and the rest of the
integrand will be independent of the $\omega$'s. Upon integrating out the
$\omega$'s and $\omtld$'s each plaquette contributes a factor
\begin{equation}
 \int\!\!d\omega_p \int\!\!d\omtld_p (\omp_p)^{n_p} (\omm_p)^{m_p} =
 \frac{1}{2} \delta_{n_p0} \, \delta_{m_p\mbox{\scriptsize even}} +
 \frac{1}{2} \delta_{m_p0} \, \delta_{n_p\mbox{\scriptsize even}},
\end{equation}
where $\delta_{n_p \mbox{\scriptsize even}}$=1 if $n$ is even, 0 otherwise.
This means that each plaquette can have only an even power of one type 
of $\omega$ (+ or -) otherwise the corresponding term vanishes.

In the following we list the expansion of different factors
in the integrand of (\ref{eq:GmG}) in terms of the $\omega^{\pm}$'s.
Since we want to verify only the positivity of each term in the expansion 
of (\ref{eq:GmG}) we shall sometimes omit positive 
constant factors and in these cases use ``$\approx$'' instead of ``=''.

\underline{The plaquette factors} can be written as
\begin{equation}
 \chi_{\omega_p}(\mbox{d}\eta_p) \, e^{h \omtld_p} =
 (\thp + \thm \omega_p) (\cosh h + \omtld_p \sinh h) \approx
 (\thp + \thm \omega_p) (1 + \omtld_p \hhat),
\end{equation}
where $\theta^{\pm}_p$ are projection operators onto states without/with
(+/-) a vortex at $p$, $\hhat= \tanh h$ and we have omitted the trivial
constant factor $\cosh h$. After the change of variables (\ref{eq:chvar})
this becomes
\begin{eqnarray}
 \chi_{\omega_p}(\mbox{d}\eta_p) \, e^{h \omtld_p}  =
 (\thm_p + \hhat \thp_p) \omp_p \; + \; (\thm_p - \hhat \thp_p) \omm_p +
      \hspace{5cm}  \nonumber \\
 \frac{1}{2}(\thp_p + \hhat \thm_p) (\omp_p)^2 \; + \; 
 \frac{1}{2}(\thp_p - \hhat \thm_p) (\omm_p)^2
     \label{eq:plaq}
\end{eqnarray}
\underline{The links not belonging to $C_2$} can be expanded as 
\begin{eqnarray}
 e^{K \sd^\star \omega_l +K \sd^\star \omtld_l} 
 \approx (1+ \khat \dstar \omega_l) (1+ \khat \dstar \omtld_l) =
       \hspace{5cm} \nonumber \\
 \khat (\dstar \omega_l + \dstar \omtld_l) \; + \; 
 \frac{1+\khat^2}{4} (\dstar \omega_l + \dstar \omtld_l)^2 \; + \;
 \frac{1-\khat^2}{4} (\dstar \omega_l - \dstar \omtld_l)^2
\end{eqnarray}
To rewrite this in terms of $\omega^{\pm}_{l1}$ and $\ompm_{l2}$,
the $\omega^\pm$ variables on the two plaquettes sharing the link $l$,
we can apply the following identies that can be easily checked by using
the definition of the $\ompm$'s;
\begin{eqnarray} 
 \omega_1 \omega_2 \pm \omtld_1 \omtld_2 & = &
 2 ( \omp_1 \ompm_2 + \omm_1 \ommp_2 )  \\
 (\omega_1 \omega_2 \pm \omtld_1 \omtld_2)^2 & = &
 4 \left[ \, (\omp_1)^2 (\ompm_2)^2 + (\omm_1)^2 (\ommp_2)^2 \, \right] + ...
\end{eqnarray}
The ellipses in the second identity means that terms that integrate to zero
because they contain both \omp\ and \omm\ on the same plaquette have been
omitted. Now the final form of the expansion for links not contained in $C_2$
is
\begin{eqnarray}
 e^{K \sd^\star \omega_l +K \sd^\star \omtld_l} =
 2 \khat (\omp_{l1} \omp_{l2} + \omm_{l1} \omm_{l2}) & + &
 (1+\khat^2) \left[ (\omp_{l1})^2 (\omp_{l2})^2 + (\omm_{l1})^2 (\omm_{l2})^2
 \right]    \hspace{5mm} \nonumber \\
 & + & (1-\khat^2) \left[ (\omp_{l1})^2 (\omm_{l2})^2 + 
 (\omm_{l1})^2 (\omp_{l2})^2 \right]
\end{eqnarray}
Finally for \underline{links contained in $C_2$} it is also straightforward 
to carry out the same type of expansion. These factors can
be of two different types depending on the corresponding sign in the 
r.h.s.\ of (\ref{eq:C2links}) carried by the link in question. 
If the sign is positive we obtain 
\begin{eqnarray}
 e^{ K (\sd^\star \omega_l - \sd^\star \omtld_l)} +  
 e^{-K (\sd^\star \omega_l - \sd^\star \omtld_l)}  & = &
 2(1-\khat^2) \left[ (\omp_{l1})^2 (\omp_{l2})^2 +
                             (\omm_{l1})^2 (\omm_{l2})^2 \right]
     \nonumber \\ 
 & & +  2(1+\khat^2) \left[ (\omp_{l1})^2 (\omm_{l2})^2 + 
                             (\omm_{l1})^2 (\omp_{l2})^2 \right]
\end{eqnarray}
For links with a minus sign the result is
\begin{equation}
 e^{ K (\sd^\star \omega_l - \sd^\star \omtld_l)} - 
 e^{-K (\sd^\star \omega_l - \sd^\star \omtld_l)} = 
 4\khat (\omp_{l1} \omm_{l2} + \omm_{l1} \omp_{l2}).
\end{equation}
We are now ready to construct the ``diagrams'' of our expansion by choosing
one term from each plaquette and link expansion in all possible ways 
consistent with the rule that eventually each plaquette has to carry
an even power of either $\omp$ or $\omm$ but not both. 
This constraint implies that the
plaquettes carrying $(\thm \pm \hhat \thp) \ompm$ are pairwise connected with
stacks of links carrying an odd power of the \ompm's. These links always
come with an additional factor $\sim \khat$ while links carrying an even
power of the $\omega$'s contain $1 \pm \khat^2$. Recall that $\khat =
\tanh K$ is small at low temperature which means that a stack of odd
$\omega$-power links costs an energy proportional to its length. Moreover
the plaquettes that are connected by these stacks carry vortices
(\thm) with a small mixing of the no-vortex state $(\sim \hhat \thp)$.
We can thus recover the familiar string vortex picture in this representation,
the only subtlety being that the effective magnetic field gives a small
mixing between \thp\ and \thm.

After integrating out the $\omega$ variables, up to positive
constant factors, each diagram has the form
\begin{equation}
 D = \langle \prod_{p \in \bar{\Lambda}} (\thm_p \pm \hhat \thp_p)
 \prod_{p \notin \bar{\Lambda}} (\thp_p \pm \hhat \thm_p) \rangle_L,
     \label{eq:vorex1}
\end{equation}
where $\langle \rangle_L$ means expectation with respect to the SO(3) measure
$d\nu_+[U]$ and $\bar{\Lambda}$ is a subset of plaquettes containing an even number of plaquette

\end{document}